\documentclass[12pt,fleqn]{article}
\usepackage{epsfig} 
\usepackage{amssymb}
\usepackage{amsthm}
\usepackage{amsmath}

\newcommand{\rem}[1]{}

\newcommand{\Eq}[1]{Eq.~(\ref{#1})}
\newcommand{\Th}[1]{Th.~\ref{#1}}
\renewcommand{\S}[1]{Sec.~\ref{#1}} 	
\newcommand{\Fig}[1]{Fig.~\ref{#1}}

\newcommand{\EpsfigStuff}[5] {
                          \begin{figure}[#2]
                              \epsfxsize=#5
                              #1 
                              \caption{\small { #3 }\label{#4}}
                          \end{figure}
}
\newcommand{\Epsfig}[5] {
                          \begin{figure}[#2]
                              \epsfxsize=#5
                              \centerline{\epsfbox{#1}}
                              \caption{\small { #3 }\label{#4}}
                          \end{figure}
}

\newcommand{\eps}{\varepsilon}
\newcommand{\teps}{\tilde\varepsilon}

\newcommand{\Res}{\mathop{\rm Res}}

\newcommand{\Imin}{I_{\rm min}}
\newcommand{\Imax}{I_{\rm max}}

\renewcommand{\d}{{\rm d}}

\newcommand{\zb} {\bar{z}}

\newcommand{\hen} {H\'enon }

\newtheorem{teo}{Theorem}
\newtheorem{lem}[teo]{Lemma}

\begin{document}

\title{Generic Twistless Bifurcations}
\author{H.~R.~Dullin, J.~D.~Meiss, D.~Sterling
\thanks{HRD was supported in part by DFG grant Du 302/2. 
	JDM was supported in part by NSF grant DMS-9623216.  
	DS was supported by an NSF Graduate Traineeship under DMS-9256335}\\
	Department of Applied Mathematics, \\
	University of Colorado, Boulder, CO 80309-0526 \\
{\small
		hdullin@Colorado.EDU,
		jdm@Colorado.EDU,
		sterling@Colorado.EDU}
}
\date{\today}
\maketitle

\begin{abstract}
We show that in the neighborhood of the tripling bifurcation of a 
periodic orbit of a Hamiltonian flow or of a fixed point of an area preserving 
map, there is generically a bifurcation that creates a ``twistless'' 
torus. At this bifurcation, the twist, which is the derivative of the 
rotation number with respect to the action, vanishes. The twistless 
torus moves outward after it is created, and eventually collides with 
the saddle-center bifurcation that creates the period three orbits. 
The existence of the twistless bifurcation is responsible for the breakdown of 
the nondegeneracy condition required in the proof of the KAM theorem 
for flows or the Moser twist theorem for maps. When the twistless torus
has a rational rotation number, there are typically reconnection 
bifurcations of periodic orbits with that rotation number.


\vspace*{1ex}

\end{abstract}

\newpage

\section{Introduction}

%
%
The dynamics in the neighborhood of an elliptic periodic orbit of a 
two-degree of freedom Hamiltonian flow, or equivalently, an elliptic 
fixed point of an area preserving map, can be elucidated by 
consideration of their formal normal forms.  When the rotation 
number, $\omega$, of the elliptic orbit is irrational, the normal form 
is called the Birkhoff normal form.  Let $J$ denote the transverse 
action (the ``symplectic radius'') and $\theta$ be its conjugate 
angle.  The Birkhoff normal form for the flow is
\begin{equation}\label{eqn:bFormH}
     H(J) = \omega J + \frac12 \tau_0 J^2 + \frac16 \tau_1 J^3 + \dots \;.
\end{equation}
For an area-preserving map with an elliptic fixed point, the Birkhoff normal 
form is
\begin{eqnarray}\label{eqn:bFormMap}
	J      &\mapsto& J  \nonumber\\
	\theta &\mapsto& \theta + 2\pi \Omega(J)
\end{eqnarray}
where the rotation number is
\[
 \Omega(J) = \omega + \tau_0J + \frac12 \tau_1J^{2} +\ldots \;.
\]
This map is also the time $2\pi$ map of the Hamiltonian flow.
We define the ``twist'' to be the derivative of the transverse rotation 
number with respect to the action:
\begin{equation}\label{eqn:twistdef}
    \tau(J) \equiv \frac{d\Omega}{dJ} = \tau_0 + \tau_1J + \ldots \;.
\end{equation}

In general, the Birkhoff series do not converge.  The actual dynamics of 
the system typically includes a chain of $n$ islands located near each 
radius where $\Omega = m/n$, a rational number.  This chain is 
constructed from an elliptic and hyperbolic pair of periodic 
orbits of period $n$.  These orbits are born in an $n$-tupling 
bifurcation that occurs when $\omega$ passes through the rational 
value $m/n$.

The character of the bifurcation as $\omega$ passes through $m/n$ when 
$\tau_0 \ne 0$ depends upon $n$ 
\cite{Ozorio88,MH92,Schomerus}.  When $n>4$, the elliptic 
orbit remains stable, and the bifurcation gives rise to a chain 
of $n$ islands.  As we recall in the Appendix, 
for $n=4$ the origin can become unstable providing the resonant term
is large enough.  When $n=3$, 
the resonant term generically dominates the twist term near the origin, and 
the origin is unstable at the bifurcation point.

When the rotation number $\Omega$ is a monotone function of $J$, or 
equivalently, the twist does not vanish, the map is called a 
``monotone twist map.'' For this case, Aubry-Mather theory applies 
\cite{Meiss92} and implies in particular that there is a pair of 
rotational $m/n$ periodic orbits for each rational in the range of 
$\Omega$.  The nonvanishing of the twist also corresponds to the 
isoenergetic nondegeneracy condition required for KAM theory 
\cite{Arnold78}, in the proof of the existence of invariant tori.

When the twist vanishes, the dynamics can be much more complicated.  
For example, ``reconnection bifurcations,'' occur near an extremal 
point of $\Omega$ \cite{HowHoh84,HowHum95}, and the renormalization 
operator for the destruction of invariant circles gives a distinct 
universality class for circles that cross the zero twist line as 
opposed to those which do not \cite{CGM96}.
Recently \cite{Simo98} it has been shown that the ``meandering curves''
that appear when a twistless curve passes through a rational rotation 
number are stable under small perturbations.
An extension of standard KAM theory \cite{DL98} shows that in two 
parameter families of area preserving maps a diophantine twistless 
curve persists.

While it appears that $\tau_0$ should generically be nonzero, and so the 
twist is nonvanishing at least in some neighborhood of the origin, we 
will show in this paper that this is not true 
whenever the rotation number $\omega$ passes through $1/3$. 
This answers the question about genericity raised in \cite{Simo98}.

%
%
To show that the twist generically vanishes, we begin with the 
normal form in the neighborhood of a tripling bifurcation.  Generally, 
when the rotation number of the elliptic periodic orbit is rational, 
there are resonant terms that cannot be transformed away.  If we keep 
only the first such resonant term, the Hamiltonian becomes (see 
e.g.~\cite{Ozorio88,MH92})
\[
  \tilde H(I,\theta,t) = \omega I + A I^2 + \dots + 
                        B I^{n/2} \cos(n\theta - m t) + \dots \;.
\]
We now use the variable $I$ to denote the ``action'' for this system; 
it is not a true action variable.
Here the system is at resonance when $\omega = m/n$. 
The time dependence in $\tilde H$ can be eliminated by 
a transformation to rotating coordinates, defining $\phi = \theta - 
mt/n$. This gives the new Hamiltonian $H = \tilde H - mI/n$:
\begin{equation}\label{eqn:rFormH}
  H(I,\phi) = \eps I + A I^2 + \dots + B I^{n/2} 
              \cos(n\phi) + \dots,
\end{equation}
where $\eps = \omega - m/n$ measures the frequency difference from the 
resonant case.  The time $2\pi$ map of $\tilde H$ is an area 
preserving map which has $H$ as a conserved quantity.

When the resonant coefficient $B$ vanishes, then the coordinate $I$ is 
the true action, $J$, and the twist for \Eq{eqn:rFormH} is $\tau = 2A+ 
O(J)$, as we can see by comparison with \Eq{eqn:bFormH}.  However, 
when $B \ne 0$, the action is modified and so is the twist.  Moreover 
when $n=3$, the resonant term is of lower order than the first twist 
term, and even the $O(J^{0})$ terms in the twist must be corrected.  
We will compute the twist for this system in \S{sec:rHam}.

We also consider the resonant normal form for an area-preserving 
map in cartesian coordinates $(p,q)$. 
This is most easily written in complex coordinates,
\begin{equation}\label{eqn:zDef}
       z \equiv \sqrt{2I} e^{2\pi i \theta} = p + iq\;,
\end{equation}
and in terms of the multiplier of the fixed point
\begin{equation}\label{eqn:lamDef}
       \lambda \equiv e^{2\pi i \omega} \;.
\end{equation}
Then the normal form can be written
\begin{equation}\label{eqn:rFormMap}
	z' = \lambda( z + i\alpha z^{2}\zb + \dots + \beta \zb^{n-1}+\dots ) \;.
\end{equation}
where the omitted terms, as we will see below, include terms that are 
required for the map to be area-preserving.  By comparison with 
\Eq{eqn:bFormMap}, the twist, when the resonant terms vanish is $\tau = 
\alpha/\pi + O(J)$.  The resonant terms, however, will modify the 
twist; indeed when $n=3$, the resonant terms are of lower order 
than the first twist term, and so even the $O(J^{0})$ terms in $\tau$ 
should be changed.

We will show that whenever the resonant term is nonzero, the twist 
vanishes in the neighborhood of the tripling bifurcation.  We will do 
this by assuming that $\eps = \omega - m/n \ne 0$ in \Eq{eqn:rFormH} and 
\Eq{eqn:rFormMap}, so that they can be transformed to Birkhoff normal 
form.  This gives an expression for the twist that diverges at 
resonance.  The twist, however, is well defined away from resonance, 
and we will see that $\tau_{0} = 0$ at some $\omega = \omega_{0}$ in the 
neighborhood of $1/3$.  At this rotation number a twistless torus is 
created at the origin.  As $\omega$ moves away from $\omega_{0}$, the 
twistless torus grows, corresponding to an extremum in $\Omega(J)$.

As Moser showed \cite{Moser94}, the twist for the \hen map vanishes at 
$\omega_{0} \approx 0.29$.  We show in addition that the twistless 
circle moves away from the origin as $\omega$ moves towards $1/3$.  
This implies, for example, that when the rotation number of the 
twistless circle passes through a low order rational number, such as $3/10$, a 
reconnection bifurcation \cite{HowHoh84,HowHum95} should occur.  The 
dynamical consequences of this have already been observed in 
\cite{WVCP88,SimTre98},
though without explaining their genericity.  Using our 
calculation of the higher order twist, we will obtain a good 
approximation for the position of this bifurcation.

More generally, the twist can vanish at any rotation number.  For 
example, for a cubic map, we will show that with the choice 
of two parameters we can make $\tau_{0}$ vanish at 
$\omega = 1/5$ or $2/5$, which results in the instability of the 
elliptic point at the bifurcation. Similarly the seventh order
resonance can be generically destabilized in three parameter family
of quartic maps. The instability of these resonant twistless maps is proved
in the appendix.

\section{Resonant Hamiltonian flows} \label{sec:rHam}
In this section we will take advantage of the fact that the resonant 
normal form \Eq{eqn:rFormH} is integrable, to obtain exact 
expressions for the twist near resonance. We begin by rewriting 
\Eq{eqn:rFormH} in the form
\begin{eqnarray}\label{eqn:rHam}
	H(I,\phi) & = & P(I) + Q(I) \cos(n\phi), \quad \mbox{with} \nonumber \\ 
	P(I) & = & \eps I + A I^2 + \dots \\
	Q(I) & = & I^{n/2}\left(B + \dots \right), \quad n > 2,
		\nonumber
\end{eqnarray}
where the dots denote finite polynomials in $I$.
Our goal is to transform $I$ to the true action variable,
$
    J \equiv \frac{1}{2\pi} \oint I \d \phi
$
for \Eq{eqn:rHam} and obtain the expression for the rotation number 
$\Omega(J)$.  To do this, we assume explicitly that $\eps \ne 0$ so 
that $P$ has a first order zero at the origin.  

To find the period, $T(h)$, we solve
$H(I,\phi)=h$ for $\phi$ and substitute this into the differential 
equation for $I$ to obtain
\[
     T(h) = \oint \frac{\d I}{\sqrt{Q^2 - (h-P)^2}}.
\]
Here we have taken into account that the extrema in $I$ are visited 
$n$ times for a full turn of $\phi$; therefore the integral has been 
multiplied by $n$.

The action is given by the area under the curve $I(h,\phi)$, but it is
better to obtain it by integrating $-\phi(I,h)\d I$, because otherwise we 
would need to solve a quartic or higher equation for $I$.  The area 
under $I$ is given by the difference between the maximum area $2\pi 
\Imax$ and $2n$ times the area under the curve $\phi$ between 
extrema, since the cosine is even, and each extremal is visited $n$ times 
in one loop.  Hence the action turns out to be
\begin{equation} \label{eqn:Jh}
	J(h) = \frac{1}{2\pi}\left( 2\pi \Imax(h) - 
		2n\int_{\Imin(h)}^{\Imax(h)}\phi(I,h) \d I \right).
\end{equation}
As usual one can verify that $T(h) = 2\pi J'(h)$, since the boundary 
terms cancel.

As expected, the period is an Abelian integral of the first kind
because $P(I)$ and $Q(I)^2$ are polynomials in $I$. The action 
can also be turned into an Abelian integral by partial integration
of Eq.~(\ref{eqn:Jh}).
Again the boundary term cancels and we obtain

\begin{lem}
The action $J(h)$ and the period $T(h)$ of the resonant normal
form Hamiltonian \Eq{eqn:rHam} are Abelian integrals
on the Riemann surface
\[
	\Gamma: y^2 = R(z) =  Q(z)^2 - (h-P(z))^2.
\]
They are given by
\begin{eqnarray}  \label{eqn:action}   
    2\pi J(h) &=& \oint_\gamma\frac{P'(z)Q(z) + Q'(z)(h-P(z))}{Q(z) y} z \d z \nonumber\\
         T(h) &=& \oint_\gamma \frac{1}{y} \d z,
\end{eqnarray}
where $\gamma$ corresponds to the real cycles of $\Gamma$.
\end{lem}

The degree of $R(z)$ is at least $n$ because the lowest order term in 
$Q(z)$ is $z^{n/2}$.  For $n=3$ the integrals are elliptic, for larger 
$n$ they are hyperelliptic.  
The equilibrium points of \Eq{eqn:rHam} correspond to the double 
roots of $R(z)$.  The values of $h$ for which this occurs are called 
critical values and they can be determined by the discriminant of $R(z)$. 
The simplest critical value is $h=0$, since there is always a double root 
at $z=0$:
\[
	R(z)|_{h=0} = Q(z)^2 - P(z)^2 = - z^2 \eps^2(1 + \dots).
\]
This corresponds to the fixed point at the origin.  Since $R''(0)<0$ 
at $h=0$ there is a maximum between the colliding roots so this 
corresponds to a vanishing real cycle $\gamma$.\footnote
	{$R'''(0) \not = 0$: for $n=3$ it vanishes at 
	$\teps = 1/2$, but then $R^{(4)}(0) - 12 < 0$.}
All that is left from the Abelian integral in this case is the residue 
of the pole at $z=0$.  The limiting period is therefore given by
\[
	T(0) = 2\pi i \Res_{z=0} \frac{1}{y} = \frac{2\pi}{\eps}
\]
This is a trivial result, because $\eps$ was designed to be
the deviation in rotation number from $m/n$ in the first place.
Note, however, that the original variable $I$ was not the
true action. It is the fact that the resonant term is of order 
$n/2$ that the frequency at $h=0$ stays the same.
A similar calculation will lead to the twist of the origin, 
which can be changed by the resonant term if $n=3$.

\subsection{Tripling bifurcation for flows}

For the tripling bifurcation the Hamiltonian from
\Eq{eqn:rFormH} through quadratic order is 
\begin{equation}\label{eqn:rFormH3}
	H(I,\phi) = \eps I + A I^2 + B I^{3/2} \cos(3\phi).
\end{equation}
Assuming that both $A$ and $B$ are nonzero, we can  
eliminate both parameters in $H$ upon defining
\[
    \tilde I = \frac{A^2}{B^2} I  \;,\quad
    \tilde h = \frac{A^3}{B^4} h  \;,\quad
    \teps = \frac{A}{B^2} \eps \;, \quad \mbox{ and } 
    \tilde \phi = \phi + \psi \;,
\]
where $\psi = 0$ if $AB>0$ and $\psi = \pi/3$ otherwise.  This scaling 
leaves only the two essential parameters $(\teps,\tilde h)$, so that 
the energy equation becomes
\begin{equation}\label{eqn:scaledEnergy}
    \tilde h = \tilde\eps \tilde I + \tilde I^{2} + \tilde I^{3/2} 
	\cos(3\tilde \phi) \;.
\end{equation}
The corresponding elliptic curve is 
\[
	R(z) =  z^3 - (\tilde h -  \teps z - z^2)^2 \;.
\]
We consider $\teps$ as the bifurcation  parameter and the energy $\tilde h$ 
is the parameter selecting a particular torus. 

\Epsfig{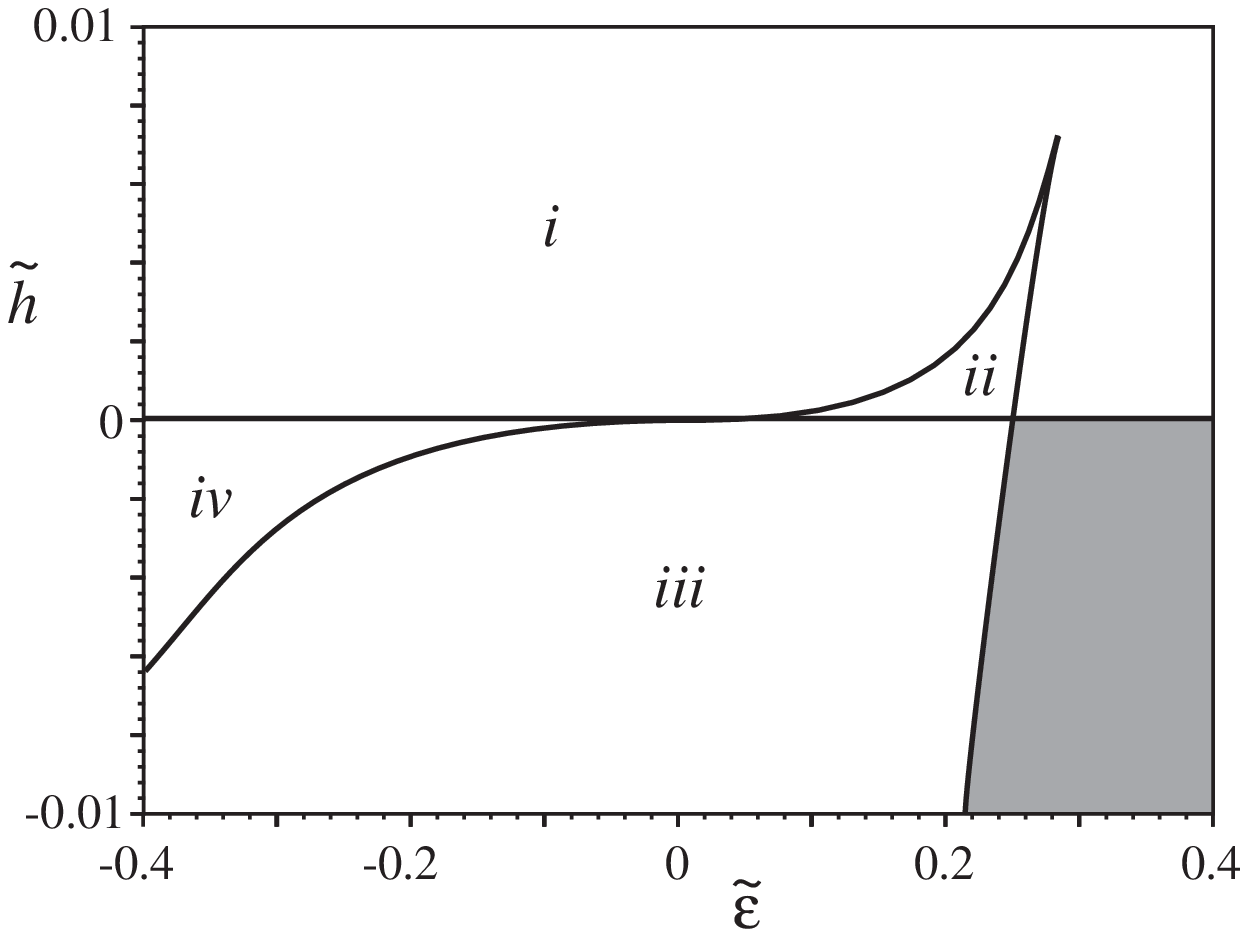}{ht}
	{Bifurcation diagram for \Eq{eqn:scaledEnergy}. The horizontal
	axis corresponds to the fixed point at the origin. The upper of
	the two remaining curves represents the saddle 3-fold orbit and the lower
	one the elliptic 3-fold orbit. At $\tilde \eps=9/32$, where the two lines meet, 
	these orbits are collide in a saddle-center bifurcation. The shaded 
	area is energetically forbidden}
{fig:bifHam3}{9cm}

The phase portrait in Cartesian coordinates, $H((p^2+q^2)/2,\arctan p/q)$, 
always has one equilibrium point at 
the origin corresponding to the line $h=0$ in \Fig{fig:bifHam3}. 
The other equilibria are easily 
found from the zeros of the discriminant $D$ of $R(z)$, which is given by
\[
	D(\tilde h, \teps) = - \tilde  h^3 \left[
             256 \tilde  h^2+ (27-144 \teps+128 \teps^{2}) \tilde  h
	         +4 \teps^{3}(-1+4 \teps) \right] \;.
\]
Apart from the triple zero at $\tilde h=0$, $D$ has two distinct real roots 
for $\tilde h$ when $\teps < 9/32$.  
These are the energy levels for one of the 3-fold 
equilibria which correspond to the period three orbits in the 
original, nonrotating frame. The point $(\teps,
\tilde h) = (9/32,(3/16)^{3})$ corresponds to the cusp in 
\Fig{fig:bifHam3}, where the period three orbits collide. There are 
four regions in the bifurcation diagram; the 
corresponding phase portraits are shown in \Fig{fig:bifscheme}. 
In region (i) there is one interval of positive $R$ corresponding to
one interval of real momenta, hence one torus. Crossing the critical 
lines to regions (ii) and (iv) creates a second interval of positive
$R$ out of a double root. Entering region (iii) from these regions 
by crossing another line of critical points destroys a positive 
interval in a double root so again there is only one torus corresponding
to each point in region (iii). Regions (i) and (iii) can also be left 
by destroying the one positive interval in a double root so that
no motion at all is possible (the shaded region).

The details of which torus in the phase portrait belongs to what
region are shown in \Fig{fig:bifscheme}.
Consider the one positive interval of region (i):
For $\teps < 0$ it corresponds to tori sufficiently far away from
	the origin (part of dotted line); 
for $0< \teps < 1/4$ it corresponds to {\em all} tori outside 
	the separatrix (dotted line); 
for $1/4 < \teps < 9/32$ it corresponds to {\em all} tori outside
	the separatrix (dotted line) and also to tori sufficiently
	close to the origin (part of solid line);
for $\teps > 9/32$ it corresponds to all tori (dotted line).
For the other regions the correspondance is analogous.
\rem{
In region (iii) there is again one positive interval for $R$.
For $\teps < 0$ it corresponds to {\em all} tori surrounding the 
	threefold elliptic point (dashed line); 
for $\teps > 0$ to those sufficiently close to the elliptic point
	(part of dashed line).
In regions (ii) and (iv) there are two different positive intervals.
In (iv) they correspond to 
	all tori in the central triangle (solid line)
	and to the tori outside but close to the separatrix 
		(part of dotted line).
In (ii) for $\teps < 1/4$ they correspond to
	{\em all} tori in the central triangle (solid line) 
	and to tori in the threefold island but sufficiently close 
		to the saddle (part of dashed line);
for $\teps > 1/4$ they correspond to 
	{\em all} tori in the threefold island (dashed line)
	and to tori in the central triangle but sufficiently close
		to the saddle (part of solid line).
}
The boundaries of the 
regions, where two roots of $R$ coalesce, correspond to the 
equilibrium solutions.  Note that the origin is a minimum of 
$H((q^2+p^2)/2,\arctan p/q)$ 
for positive $\teps$ and a maximum for negative $\teps$, showing that it 
is stable when $\teps \ne 0$.

\Epsfig{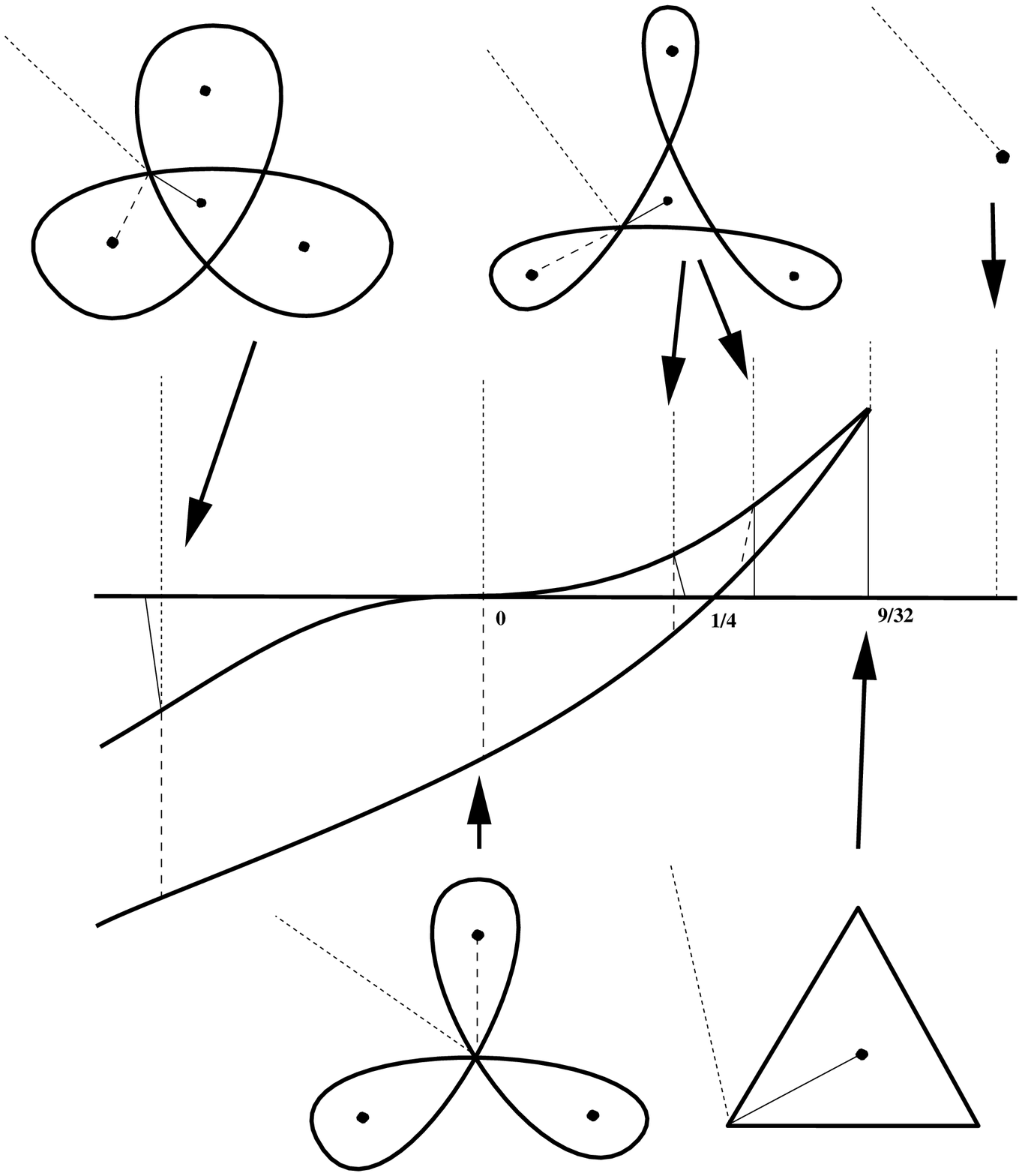}{ht}
	{Bifurcation diagram for the Tripling Bifurcation in 
	$( \teps, \tilde  h)$.  The 
	thick curves represent the energy levels of the equilibrium points.
	The phase portraits correspond to fixed values of $\teps$, and 
	the thin, vertical lines represent the energy ranges for the  
	topologically distinct families of tori, shown in each phase portrait.
	Each  group of thin lines corresponds to a single $\teps$, 
	but they are drawn slightly skewed for clarity.}
{fig:bifscheme}{7cm}

\subsection{The twist}\label{sec:HamTwist}

The twist $\tau(h)$ of a torus with energy $h$ is given by 
\Eq{eqn:twistdef}, thus
\[
	\tau(h) = \frac{\partial \Omega}{\partial J} = 
	   \frac{\Omega'(h)}{J'(h)} = -J''(h)/J'(h)^{3}.
\]
or, expanding near the origin in a series in $J$, we have
\[
    \tau(J) = -\eps^{3} J''(0) + 
                \eps^{4} \left[ 3\eps J''(0)^{2}-J'''(0) \right] J +O(J^{2})
\]
To compute these first two terms in the twist, we must find the 
second and third derivatives of $J(h)$ at $h=0$.

\begin{lem}\label{thm:hamtwist}
The twist of the fixed point at the origin of the 
Hamiltonian \Eq{eqn:rFormH3} is given by \Eq{eqn:twistdef} where
\begin{eqnarray}\label{eqn:htwist}
	\tau_{0} &=& 2A - \frac{3B^2}{2\eps}  \nonumber\\
    \tau_{1} &=& \frac32 \frac{B^2}{\eps^3} ( 8\eps A - 3B^2),
\end{eqnarray}
\end{lem}

\proof 
Upon manipulation of \Eq{eqn:action}, we can reduce the integrals to calculate 
to
\begin{eqnarray*}
	J''(h)  &=& \frac{1}{2\pi} \oint \frac{h-P(z)}{y^3} dz \\
	J'''(h) &=& \frac{1}{2\pi} \oint \left(\frac{3Q^{2}}{y^{5}} -\frac{2}{y^{3}} \right) dz
\end{eqnarray*}
To evaluate the twist we need to compute these integrals at $h=0$.  
Since the cycles reduce to loops around the origin, this amounts to 
computing residues of the integrands. 
\qed

A similar calculation for $n>3$ in \Eq{eqn:rFormH}, shows that the lowest order 
twist $\tau_{0}$ is independent of $B$.  Thus it vanishes only if the 
bare twist term, $A = 0$.  This shows that the tripling normal 
form is the only one for which the twistless torus is generically created 
at the origin.  The reason that the tripling is different is that the order 
of $Q(I)$ is smaller than that of the twist term $A I^2$.  

Lemma~\ref{thm:hamtwist} implies that the twist vanishes at the origin when 
\begin{equation}\label{eqn:epszero}
     \eps = \eps_0 \equiv \frac{3B^{2}}{4A}  \;,
\end{equation}
which corresponds to the frequency, in the nonrotating frame, of
\[
     \omega_{0} =  \frac13 +\frac{3B^{2}}{4A} \;.
\]
The twistless curve moves away from the origin as $\eps$ moves 
towards zero, corresponding to approaching the tripling bifurcation.
Using these expressions, we can compute the rotation number of the 
twistless curve near $\eps_{0}$ where it is created. In the 
original nonrotating frame, the rotation number is given by
\[
     \Omega_{0}(\omega) = \omega - \frac13 
              \frac{A}{B^2} (\omega - \omega_{0})^{2} +O(\omega - \omega_{0})^{3}
\]
where we have expressed it as a function of the rotation number 
$\omega$ of the origin.  The twistless torus can be seen in 
\Fig{fig:bif3wind}, which is a contour plot of the rotation number on 
the bifurcation diagram.  Since $\Omega'(h) = 0$ at the twistless 
torus, it corresponds to the locus of points where the contours are 
vertical.  The twistless torus emerges from the origin at $\teps=3/4$, 
and collides with the saddle-center bifurcation at $\teps=9/32$.
	In region (ii) the 
	plotted curves correspond to the rotation number of the torus in the 
	central triangle; in region (iv) they give the rotation number of the 
	torus in the outer region.  In this way the rotation number can be 
	made continuous.  The torus in the central triangle corresponding to 
	region (iv) has the negative rotation number of the one in the outside 
	region with the same $h$.  This is a result of the sum rule 
	$\oint_{\gamma_1+\gamma_{2}} \frac{\d z}{y} = 0$, where the $\gamma_i$ 
	are the two real cycles for $n=3$.

\Epsfig{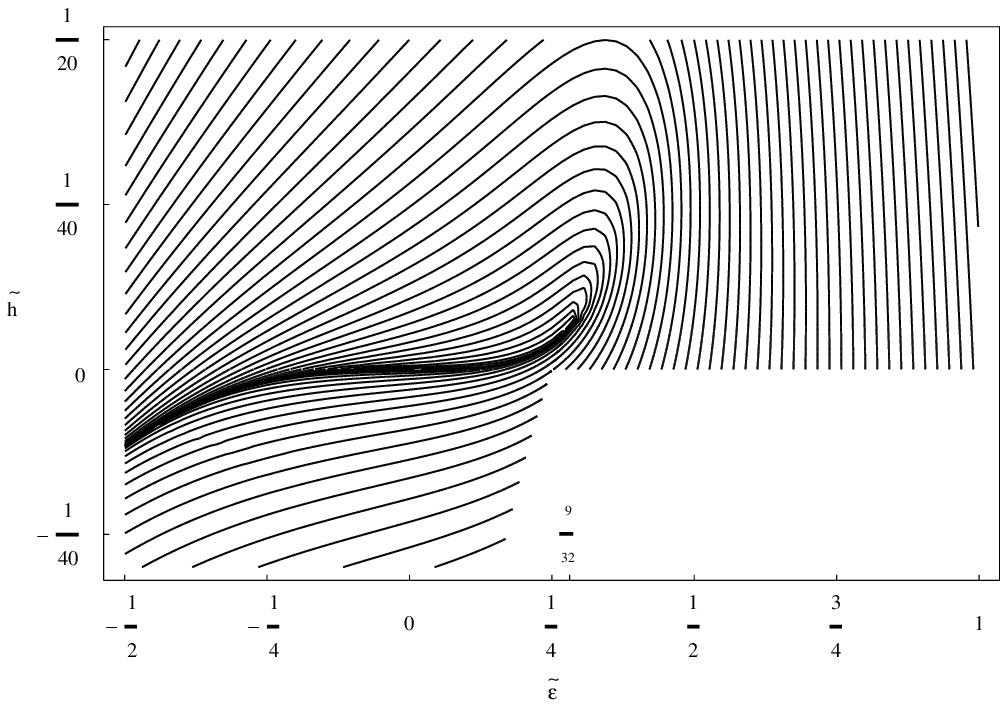}{ht}
	{Lines of constant rotation number spaced equidistant with 
	$\Delta \Omega = 0.02$
	on the bifurcation diagram for the tripling bifurcation.  
	The contours cluster 
	around the line of the unstable 3-fold orbit, because there 
	the period $T$ diverges logarithmically.}
{fig:bif3wind}{9cm}

Since the expression for $\tau_{0}$ is not affected by higher 
order terms in $P$ or $Q$, we can describe the generic scenario 
of the tripling bifurcation as follows.  A sufficient distance from 
the tripling bifurcation, when $\teps > 3/4$, the periodic orbit has 
twist, i.e., the rotation number is a monotone function of the 
transverse action near the orbit.  Note that this corresponds to 
$\omega > 1/3$ if the rotation number increases with action, i.e., 
$A>0$, otherwise it corresponds to $\omega < 1/3$.  As $\omega$ moves 
towards $1/3$ or equivalently as $\teps$ decreases toward zero, but 
before there is any obvious sign of the $1/3$ bifurcation, the central 
periodic orbit loses its twist at the parameter value $\teps = 3/4$.  

Beyond that point the twist in the island is no longer monotone, and 
there is a twistless torus somewhere in the island.  
For $\teps < 3/4$ a generic perturbation of $H$ cannot destroy a
diophantine twistless curve with nonzero second twist, as ist
shown in \cite{DL98}. The difference to ordinary KAM theory is that 
for small perturbations the twistless torus with the same rotation number 
exists not necessarily for the same but for slightly change parameter $\teps$.
At $\teps = 9/32$ the twistless torus reaches rotation 
number $1/3$ and there is a saddle-center bifurcation creating a pair of 
period three orbits at nonzero radius in the island.  
The twist inside the central triangle is 
opposite to that outside the period three island chain.  Beyond this 
point the twistless torus is replaced by the separatrix with rotation 
number $1/3$.  Of course, the separatrix is generically a 
homoclinic tangle.  

Eventually, when $\omega=1/3$ or $\teps=0$, the saddle period three 
orbit collides with the central periodic orbit.  Moving past this 
point, the saddle period three orbit re-emerges, 
and the origin has changed the sign of its twist.
Thus for $\teps < 0$, the rotation number is again a 
monotone increasing function.

We conclude by stating this more formally 
\goodbreak
\begin{teo}\label{thm:Hbif}
 Near a tripling bifurcation of a periodic orbit, a two degree of 
 freedom Hamiltonian system has a codimension one twistless 
 bifurcation, corresponding to the vanishing of the twist at the 
 periodic orbit.  For the normal form, \Eq{eqn:rFormH3}, 
 this results in 
 the creation of a twistless curve at $\teps_0 = 3/4$, \Eq{eqn:epszero}, 
 that moves away 
 from the origin as the tripling point is approached, and eventually 
 collides with the saddle-center bifurcation of the period three orbits.
\end{teo}

\section{Tripling bifurcation for area-preserving maps}
In this section we study the normal form for an area-preserving map,
$f$, in the neighborhood of an elliptic fixed point with a tripling 
bifurcation. We will show that the twist generically vanishes in the 
neighborhood of this bifurcation.

Suppose that $f$ has an elliptic fixed point at the origin with rotation 
number $\omega$, i.e., with  linear multipliers $\lambda$ and $\bar 
\lambda$, \Eq{eqn:lamDef}.  We can put the 
linearization of the map in the complex, diagonal normal form 
$(z,\zb) \mapsto (\lambda z,\bar{\lambda}\zb)$ by 
introducing a pair of complex variables $(z,\zb)$ defined by the 
linear transformation
\begin{equation}\label{eqn:xytoz}
	\left(x,y\right)^{T} = \frac{1}{2i} (vz- \bar{v}\zb)
\end{equation}
where $v$ is the complex eigenvector of $Df(0)$ associated with 
$\lambda$.  If we choose the normalization of the eigenvectors so that 
$\bar{v}\times v = 2i$, then the $z$ variables are related to 
action-angle variables by \Eq{eqn:zDef}.
In the new variables, the map has the  power series
\begin{equation} \label{eqn:series}
   z \mapsto \lambda \left[z + \sum_{n=2}^{r}\sum_{j=0}^{n} 
   a_{j,n-j}z^{j}\zb^{n-j}\right]  + O(r+1) 
\end{equation}
where the terms are ordered as homogeneous monomials of degree $n$ in 
$z$ and $\zb$.  The map for $\zb$ is simply the complex conjugate of 
\Eq{eqn:series}.  Note that the complex coefficients, $a_{jk}$, are 
not all independent when the map is area-preserving.

The normal form for \Eq{eqn:series} can be computed by applying a 
sequence of canonical transformations to eliminate as many terms in 
the series as possible \cite{SM71}. As is well known each of the coefficients 
$a_{jk}$ can be formally eliminated providing the multiplier does not 
satisfy a  ``resonance'' condition of the form
\begin{equation}\label{eqn:rescon}
        \lambda^{j-k-1} = 1
\end{equation}
In particular for any $\lambda$, the terms with $j-k = 1$ 
cannot be eliminated---these we call the ``twist'' terms, since they 
give rise to shear in the rotation about the fixed point. These terms 
give rise to the Birkhoff normal form \Eq{eqn:bFormMap}.
However, if $\lambda$ is a root of unity, then other resonant terms 
occur.  For example, when $\lambda = \frac13$, then we cannot 
eliminate terms for which $j-k = 1 \mod 3$. In this case the resonant 
normal form is
\[
    z \mapsto \lambda \left[ 
	            z + a_{02}\zb^{2}+ a_{21}z^{2}\zb + a_{13}  z \zb^{3} + 
	            a_{40}z^{4}+  a_{32} z^{3}\zb^{2} + a_{05}\zb^{5}
                    \right] +O(6)
\]

To study the dynamics in the neighborhood of the tripling bifurcation, 
we assume that the lowest order resonant term, $a_{02}$, is nonzero. In this 
case we can scale $z$ to eliminate this coefficient.
When the map is area-preserving, the thirteen real coefficients in the resonant 
normal form must satisfy four conditions through fifth order. After 
applying these conditions and scaling the map
to eliminate the coefficient $a_{02}$, we can reduce the resonant 
normal form to a seven parameter family through 5th order 
\begin{eqnarray}\label{eqn:resMap}
     z \mapsto \lambda \big[ z &+& \zb^2+ (1+i\alpha) z^2 \zb + (\beta+i\gamma)z \zb^3   \\ \nonumber
	                &+& a_{40}z^4 +  (\delta+i\eta) z^3 \zb^2 + 
	                  a_{05}\zb^5 \big] +O(6)
\end{eqnarray}
The real coefficients $\alpha, \beta, \gamma, \eta$ and the 
complex coefficient $a_{05}$ are arbitrary and 
\[
	  a_{40} = \frac14 [2+\beta+i(2\alpha-\gamma)] \; , \quad \quad
	  \delta = 2\beta -\frac{1+\alpha^{2}}{2}
\]

Taking the resonant form \Eq{eqn:resMap} as our model, we imagine that 
the coefficients are fixed, and that $\omega$ is the bifurcation 
parameter. When $\omega \ne 1/3$ the resonant form can be transformed 
to Birkhoff normal form. In fact if $\omega$ is not $1/4, 1/5$ or $2/5$,
we can reduce the map to Birkhoff normal form through 
fifth order. Our goal is to compute the twists $\tau_{0}$ and 
$\tau_{1}$ for this map.

To carry out the transformation we must apply a sequence of four, near 
identity coordinate transformations to eliminate successively all of the quadratic, 
cubic, quartic and quintic terms except for the twist terms. 
One way to obtain area-preserving coordinate transformations is 
to use canonical generating functions of the form
\[
     Pq + F^{(r)}(P,q) 
\]
where $F^{(r)}$ is homogeneous of degree $r+1$.  This implicitly 
generates a canonical transformation through the equations
\[
      p = P + \frac{\partial F^{(r)}}{\partial q}(P,q)  \quad , \quad
      Q = q + \frac{\partial F^{(r)}}{\partial P}(P,q)
\]
The lowest order terms of the transformation $(q,p) \mapsto (Q,P)$ 
are easily obtained explicitly
\begin{equation} \label{eqn:ctrans}
      P = p - \frac{\partial F^{(r)}}{\partial q}(p,q)  
         + O(2r-1), \quad
      Q = q  + \frac{\partial F^{(r)}}{\partial p}(p,q)
         + O(2r-1) \;,
\end{equation} 
and the higher order terms can be obtained with more effort, 
order-by-order.  Converting the transformation to complex 
coordinates, with $Z=P+iQ$, yields a transformation of the form
\[
   Z = z + \sum_{i=0}^{r} b_{i,r-i}z^{i}\zb^{r-i} + O(2r-1)
\]
where the coefficients $b_{ij}$ can be explicitly computed in terms 
of the coefficients of $F^{(r)}$.

We wish to compute the nonresonant Birkhoff normal form for 
\Eq{eqn:resMap} through fifth order.  To do this, we first apply the 
canonical transformation $F^{(2)}$, with the coefficients chosen to 
eliminate the quadratic term in \Eq{eqn:resMap}.  This is possible 
whenever $\omega \ne 1/3$.  This transformation generates terms of 
cubic and higher order in the map.  Whenever $\omega \ne 1/4$ we can 
then apply a transformation generated by $F^{(3)}$ to eliminate all of 
the cubic terms, except for the twist term 
$z^{3}\zb^{2}$.  This transformation will not modify the quartic terms 
in the map, but because of the $O(5)$ correction terms in 
\Eq{eqn:ctrans}, it will modify the quintic terms.  Finally, whenever 
$\omega \ne 1/5$ or $2/5$, we can apply transformations 
$F^{(4)}$ and $F^{(5)}$ to eliminate all of the quartic and all of the 
quintic terms except for the twist $z^{3}\zb^{2}$.  
These transformations will not modify the coefficient of the twist 
term, however, and therefore we do not need to compute them 
in order to find the Birkhoff normal form to fifth order.
This yields 
the Birkhoff normal form for \Eq{eqn:resMap} in the form of
\Eq{eqn:bFormMap} where
\begin{eqnarray}\label{eqn:rMapTwist}
	\pi\tau_0  &=& \alpha - \frac{(3t^{2}-1)}{t(t^{2}-3)}  \nonumber\\
	\pi \tau_1 &=& 4(\eta-\gamma) + 
	           12\frac{3t^2-1}{t(t^2-3)} 
	                \left(\frac{(t^{2}+1)^{3}}{2t^2(t^2-3)^2}
                          + \pi\tau_0 \frac{3t^2-1}{t(t^2-3)} -\beta 
                    \right)
\end{eqnarray}
where $t \equiv \tan(\pi \omega)$. 

Near $\eps = 0$ we can reduce these expressions to
\begin{eqnarray}\label{eqn:rMapT2}
 \pi\tau_0 &=&  - \frac{1}{3\pi\eps} +\alpha + O(\eps) \nonumber\\
 \pi\tau_1 &=& \frac{2}{9\pi^3 \eps^3}+ \frac{4}{3\pi\eps^{2}} \tau_0
               -\frac{4}{\pi\eps}\beta + 4(\eta-\gamma -2\tau_{0}) + O(\eps^{1})
\end{eqnarray}
The dominant terms in these expressions, though they look quite 
different, are actually equivalent to those obtained for the flow in 
Lemma~\ref{thm:hamtwist}%
\footnote{
    The time $2\pi$ map of the flow of \Eq{eqn:rFormH3}, is approximately
\[
  \zeta'  \mapsto \lambda \left(\zeta + 2\pi i A |\zeta|^2 \zeta + 
\frac{3\pi i B}{\sqrt{2}} \bar \zeta^2 \right)
\]
    which is equivalent to \Eq{eqn:resMap}, upon identifying $\alpha = 
    \frac{2\pi A}{s}$, where $s = (3\pi B)^{2}/2$ is the factor that scales 
    the action to normalize $a_{02}$ in the map.}
providing we set $\beta=\eta=\gamma=0$. Nonzero values 
would correspond to higher order terms in the flow Hamiltonian.

The function $\tau_{0}(\omega)$ in \Eq{eqn:rMapTwist} is shown in 
\Fig{fig:nftwist}.
It is monotone increasing and maps the 
domain $[\frac16,\frac13) \cup (\frac13,\frac12)$ one-to-one onto 
$(-\infty,\infty)$.  Moreover, the coordinate transformations that we 
carried out to compute $\tau_{0}$ are valid in this domain.  Thus for 
any $\alpha$, there is a rotation number $\omega_{0}$ for which 
$\tau_{0}(\omega_{0}) = 0$.  If $\omega_{0}$ is not $1/5,1/4$ or $2/5$ 
then the transformations leading to the expression for $\tau_{1}$ are 
valid.  Whenever $\tau_{1}(\omega_{0}) \ne 0$, as is generically true 
for our expressions, the Moser twist theorem \cite{SM71} implies that 
there are invariant circles in the neighborhood of the elliptic point.
Thus we can conclude
\begin{teo}
    Let $f_{\omega}$ be area-preserving map with an elliptic fixed point 
    that has rotation number $\omega$.  Suppose that when $f$ is transformed to 
    its resonant normal form, \Eq{eqn:resMap}, the coefficient $\alpha$ is 
    finite (i.e.  that $a_{02} \ne 0$).  Then there is an $\omega_{0} \in 
    [\frac16,\frac13) \cup (\frac13,\frac12)$ such that $f_{\omega_{0}}$ 
    has a twistless bifurcation, i.e.  where $\tau_0(\omega_{0}) = 0$, 
    from \Eq{eqn:rMapTwist}.  The elliptic fixed point is stable whenever 
    $\omega_{0} \not\in \{1/5,1/4,2/5\}$, and $\tau_{1}(\omega_{0}) \ne 
    0$, as is generically the case.
  \end{teo}
\Epsfig{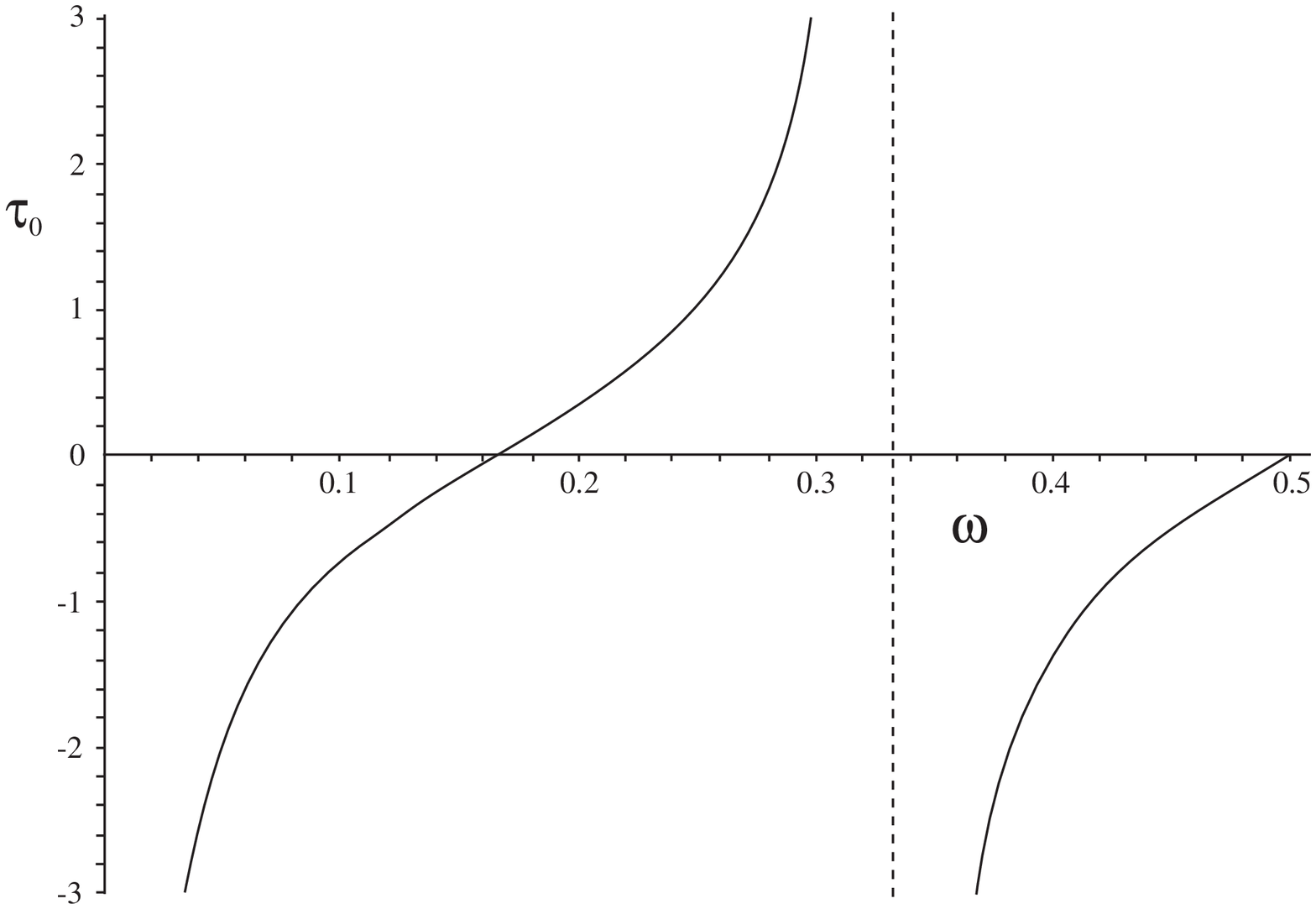}{tbh}
	{Plot of the twist $\tau_0(\omega)$ 
	 in \Eq{eqn:rMapTwist} with $\alpha=0$.}
{fig:nftwist}{9cm}

Note that in \Eq{eqn:rMapT2} the dominant contribution to 
$\tau_{1}(\omega_{0})$, when $\omega_{0}$ is reasonably close to 
$\frac13$, has the sign of $\epsilon$.  Thus there are two typical 
scenarios for the twistless bifurcation.  If $\omega_{0}< \frac13$, 
then typically $\tau_{1}(\omega_0) <0$, 
and so the rotation number $\Omega(J)$ is 
locally monotone decreasing as $\omega \rightarrow \omega_{0}$ from 
below.  The twistless curve is created at a local maximum of 
$\Omega(J)$ which moves away from the origin as $\omega$ approaches 
$\frac13$.  Conversely, if $\omega_{0}>\frac13$, then typically, 
$\tau_{1}(\omega_{0}) > 0$, so the rotation number is monotone 
increasing as $\omega \rightarrow \omega_{0}$ from above.  The 
twistless curve corresponds to a local minimum of $\Omega(J)$ which 
moves away from the origin as $\omega$ moves towards $\frac13$.  On 
the other hand, it is also possible that $\tau_{1}(\omega_{0})$ has 
the opposite sign of $\epsilon$, in which case the bifurcation creates 
a twistless curve as $\omega$ moves from $\omega_{0}$ in the direction 
away from $\frac13$.

In the actual dynamics we expect that when the rotation number of the 
twistless curve passes through rational points, there will be 
reconnection bifurcations \cite{HowHoh84,HowHum95} in which two island 
chains with the same rotation number annihilate each when they collide 
at the twistless ``curve.''  We expect this to happen 
often for the island chains with rotation numbers in the interval 
$[\omega_{0},\frac13)$, because after birth, they will travel 
outward from the fixed point, and if they move rapidly enough
will cross the twistless curve.

In the next sections, we will give examples of the twistless bifurcation
for the \hen map, and for a quartic map, where we can choose parameters 
so that $\omega_{0}$ is a low order rational.

\subsection{Tripling bifurcation for the \hen map}

Any quadratic, area-preserving map of the plane can be written in the form
\cite{Henon69}
\[
(x,y) \mapsto (y-k+x^{2},-x) \;.
\]
It has an elliptic fixed point at $x_{e} = -y_{e} 
=1-\sqrt{1+k}$ when $-1 < k < 3$. The rotation number at the fixed 
point is
\[
   \omega = \frac{1}{\pi} \arcsin\left(\frac{1+k}{4}\right)^{1/4} \;.
\]
The normalized, complex eigenvector is $v = 
\frac{1}{\sqrt{s}}(-\lambda,1)^{T}$, where $s = \sin(2\pi\omega)$.  
Using \Eq{eqn:xytoz}, shifted to the elliptic fixed point, we 
define the complex variable
\[
z = -\frac{1}{\sqrt{s}} \left(x-x_{e}+\bar{\lambda}(y-y_{e})\right)  \;,
\]
so that the \hen map becomes
\[
   z \mapsto \lambda z + \frac{1}{4s^{3/2}} 
                \left( -\lambda z+ \bar{\lambda}\zb \right)^{2} \;.
\]
Applying the normal form transformation reduces the map to the 
Birkhoff normal form, \Eq{eqn:bFormMap}, where
%
\begin{eqnarray*}
   \tau_0 &=& \frac{1}{2^{6}\pi}{\frac {(3\,t^2-5)(1+t^{2})^{3}}{t^{4}(3-t^{2})}} \;,\\
   \tau_1 &=& \frac{1}{2^{13}\pi}{\frac {(1+t^{2})^{6}
           (51t^{10}-637t^{8}+2038t^{6}-2706t^{4}+2055t^{2}-705)}
             {t^{9}(t^{2}-1)(t^{2}-3)^{3}}}  \;.
\end{eqnarray*}
%
%
where $t=\tan(\pi\omega)$, as before.
Moser previously obtained $\tau_0$ in 
\cite{Moser94}\footnote{
   Moser's equivalent expression is apparently not written in standard 
   canonical variables, and differs from ours by a scaling factor (of $-1/2$)}.
Note that $\tau_0$ vanishes only when $t=\sqrt{5/3}$, which occurs 
when $\omega = \omega_{0} = \frac{1}{2\pi}\arccos(-\frac14)) \approx 
0.2902153116$, or $k=9/16$.  As Moser noted, $\omega_{0}$ is a 
transcendental number, and therefore satisfies a Diophantine 
condition.  According to a theorem of R\"{u}ssmann, this implies that 
there are invariant curves in the neighborhood of the fixed point so 
it is stable.  This result is obtained more easily by noting that 
$\tau_1(\omega_{0}) \approx -.3366$ is nonzero.  In this case the 
stability of the fixed point follows from the Moser twist theorem 
\cite{SM71}.

\Epsfig{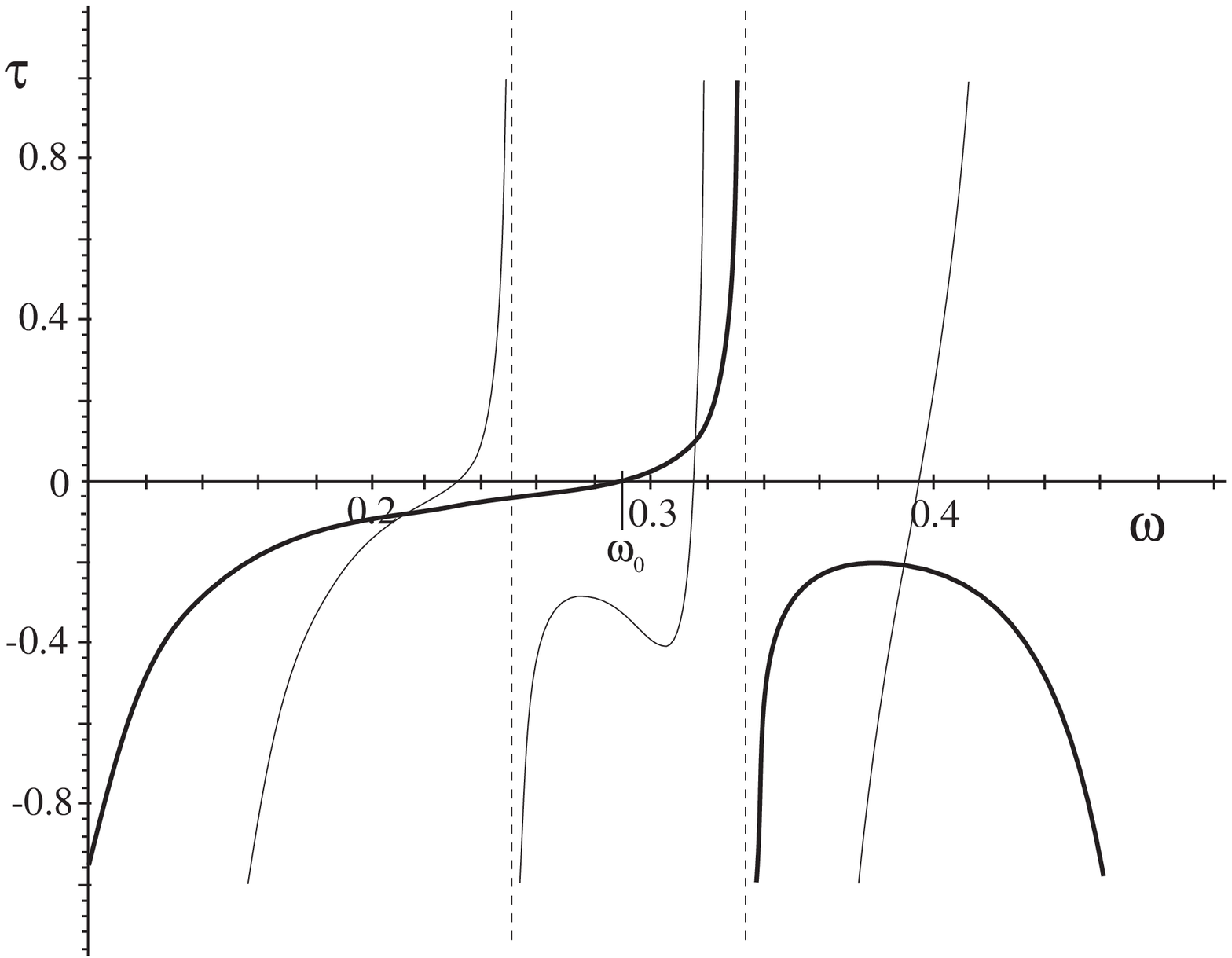}{ht}
	{Plot of $\tau_0$ (thick line) and $\tau_1$ (thin line) for the \hen 
	map as a function of $\omega$.  Here $\tau_0$ vanishes only at $\omega 
	\approx 0.2902153116$, and $\tau_1$ at $\omega \approx 0.2308206101$, 
	$0.3137515644$, and $0.3944381765$.}
{fig:htwist}{8.8cm}

Both twist functions are shown in \Fig{fig:htwist} as a 
function of $\omega$.  The Birkhoff normal form has a curve of zero 
twist that is born at $\omega_{0}$ and moves outward in action as 
$\omega$ increases at least up to $0.31375$ where $\tau_1$ vanishes.  
This upper endpoint is artificial, however, since the higher order 
twists will become increasingly important as $\tau_{1}$ nears zero.  
The rotation number of the twistless curve is approximately
\begin{equation}\label{eqn:Omega}
   \Omega_{0}(\omega) \approx \omega -\frac{1}{2}\frac{\tau_0^2}{\tau_1}
\end{equation}
This function is shown in \Fig{fig:hzerot}.  Based on the flow results,
we expect that the 
twistless curve in some sense collides with the saddle-center point of 
the period three orbit, $k=1$---this corresponds to $\omega = \omega_{5}$ in the 
figure, so that $\Omega_{0}(\omega_{5}) = 1/3$.  We saw 
in \S{sec:rHam} that the twistless curve does collide with the 
saddle-center bifurcation for the resonant normal form of the flow, which was 
integrable and for which we obtained an exact expression for $\Omega$.  
The predictions of the normal form for the map, however, are not of 
much use here because the \hen map exhibits considerable chaos for 
orbits at these radii.  If the twistless ``curve'' continues to the 
tripling, it will no longer be an invariant circle; presumably it will 
be a cantorus.  

\Epsfig{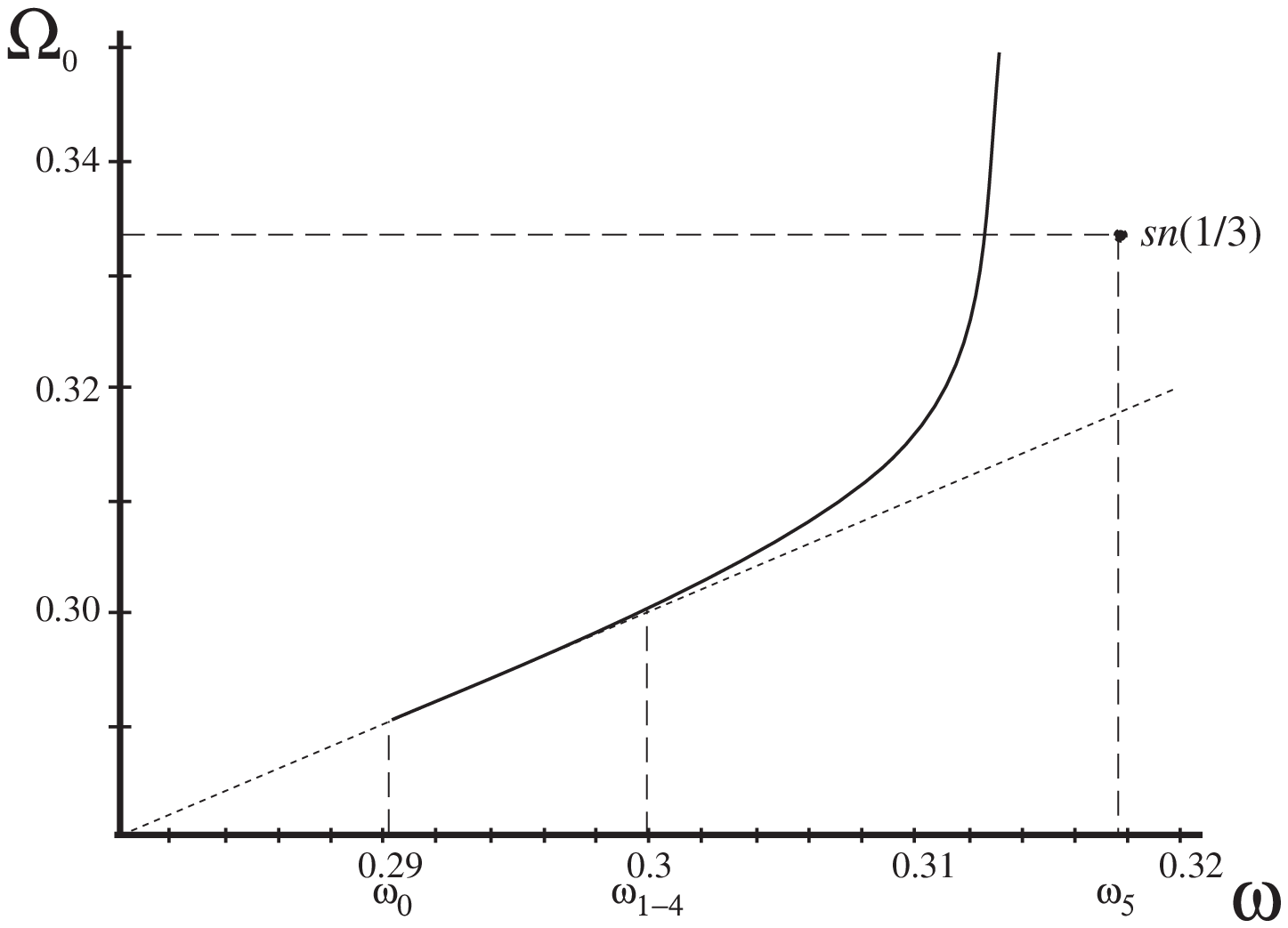}{ht}
{The rotation number of the twistless circle, $\Omega_{0}(\omega)$ near the 
tripling bifurcation of the \hen map.}
{fig:hzerot}{9cm}

In the interval where the twistless curve exists, $\omega_{0} < \omega 
< \omega_{5}$, there should be many unusual bifurcations.  For 
example, whenever the twistless curve crosses a rational rotation 
number, there will typically be a collision of two island chains of 
periodic orbits with this rotation number.  The most prominent case 
corresponds to the lowest order rational in this interval, $3/10$, 
which we sketch in the phase portraits in \Fig{fig:henonMap}.  
Similar bifurcations can be observed when the twistless curve 
passes through e.g.~$4/13$ or $5/16$.
The normal form indicates 
that the twistless curve has rotation number $\Omega_{0}(\omega) = 
3/10$, when $\omega \approx 0.2995198$, using the expansion 
\Eq{eqn:Omega}.  Translating back to the parameter of the \hen map, 
this corresponds to $k \approx 0.7060175$.  Indeed there is a 
saddle-center $3/10$ bifurcation near $k = 0.7063832$, which differs 
from our prediction by $0.05\%$.  However, the situation is more 
complicated than the normal form would indicate.  We 
list the bifurcations in Table~\ref{tbl:henbifs}, and sketch
the corresponding situation in \Fig{fig:bifwithten}.  The four $3/10$ 
orbits are created in two successive $3/10$ saddle-center bifurcations 
at $\omega_{1}$ and $\omega_{2}$.  In the interval $\omega_{2}<\omega 
<\omega_{4}$ there are four $3/10$ orbits, two are elliptic and two 
are hyperbolic.  The stable and unstable manifolds of the two 
hyperbolic $3/10$ orbits undergo a reconnection bifurcation 
\cite{HowHum95} near $\omega_{3}$, see \Fig{fig:henonMap}.  
Subsequently the ``inner'' pair of $3/10$ orbits collides with the 
elliptic fixed point, in a decupling bifurcation at $\omega_{4} = 
3/10$.  The outer pair of $3/10$ orbits move away from the origin and 
persist as $k\rightarrow \infty$.  These last two orbits are the 
orbits that we continue from the ``anti-integrable'' limit, 
\cite{SDM98}.

\EpsfigStuff{
\centerline{
\epsfxsize=4.6cm\epsfbox{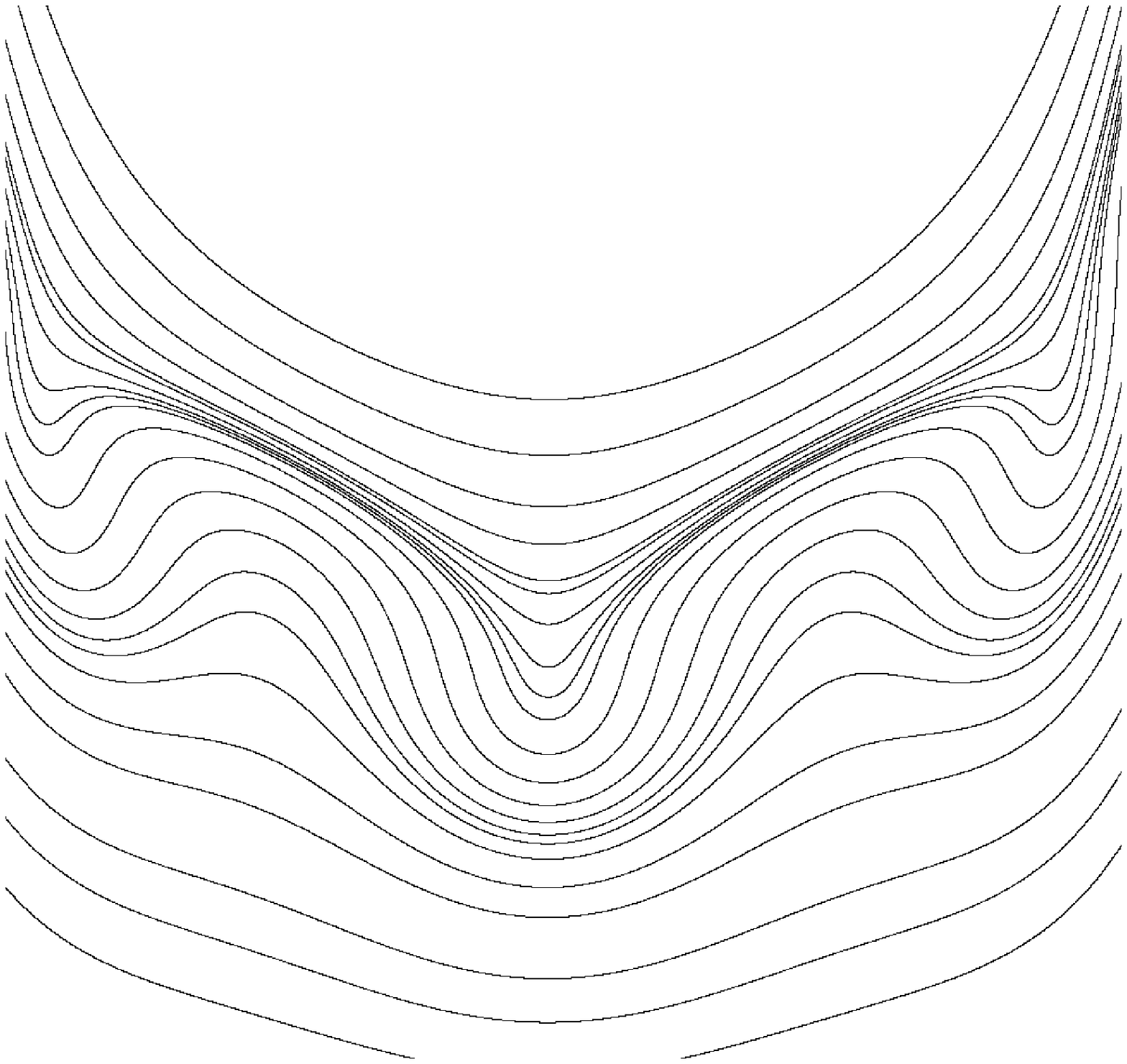}
\epsfxsize=4.6cm\epsfbox{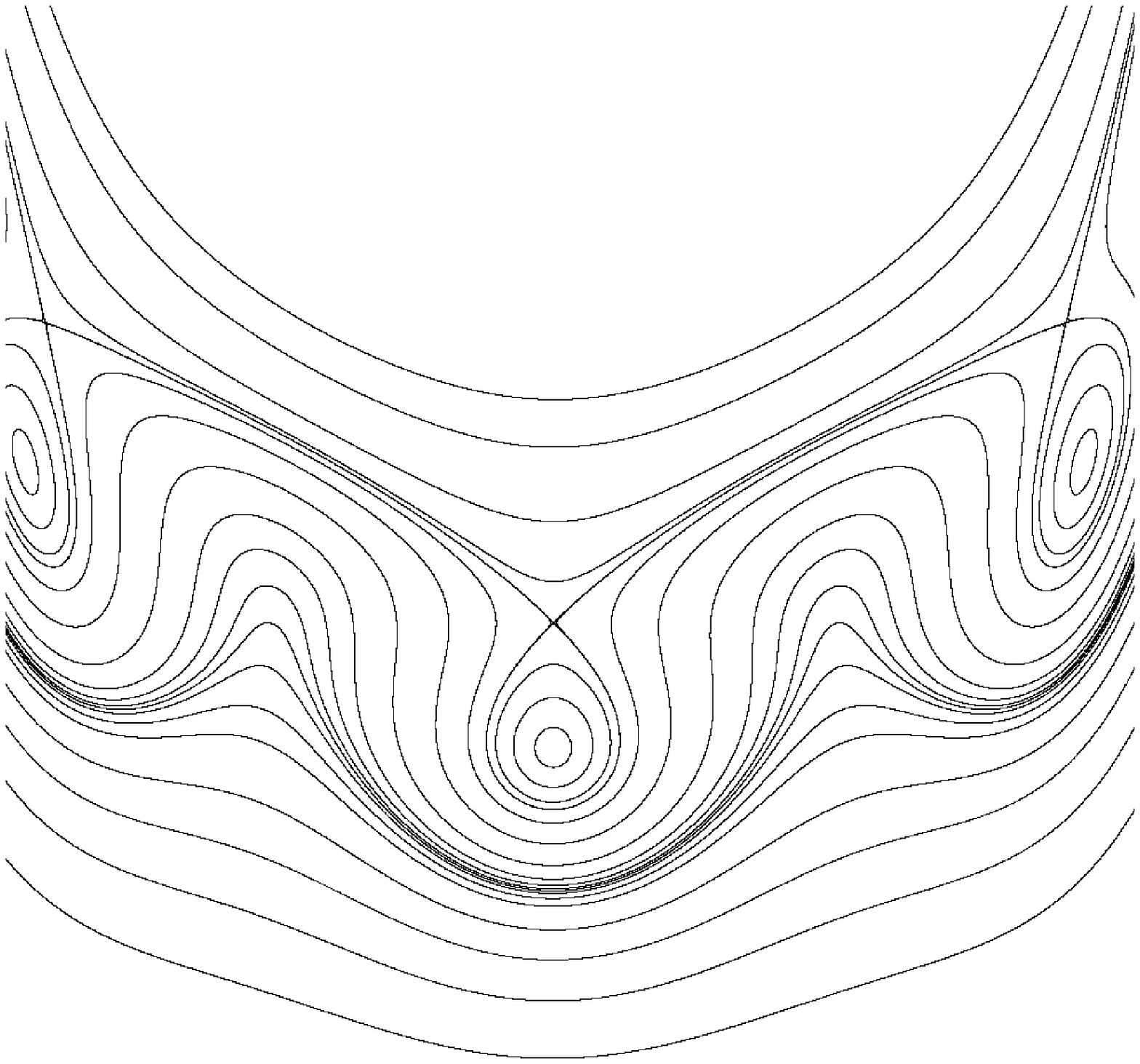}
\epsfxsize=4.6cm\epsfbox{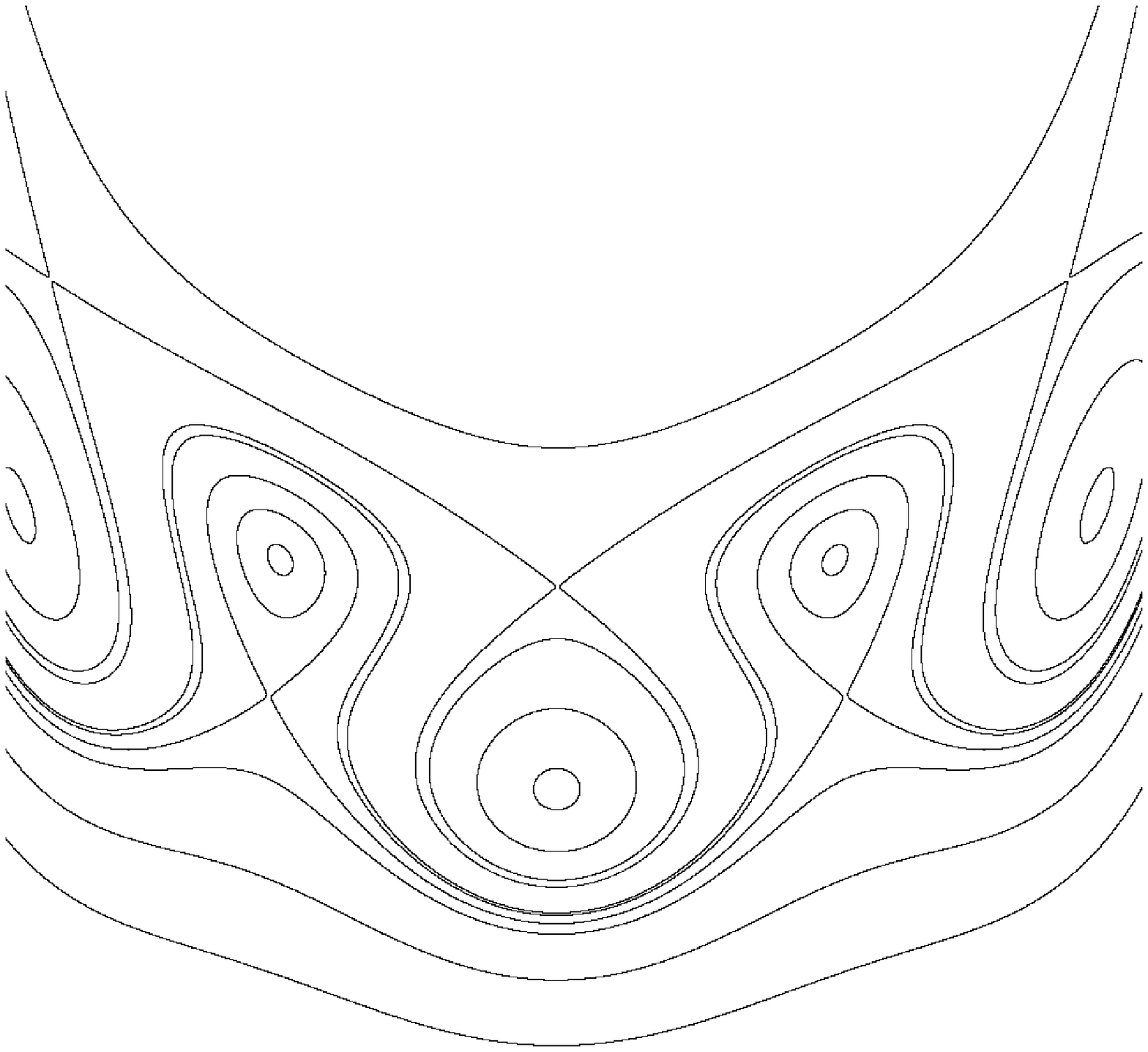}
}
\centerline{
\epsfxsize=4.6cm\epsfbox{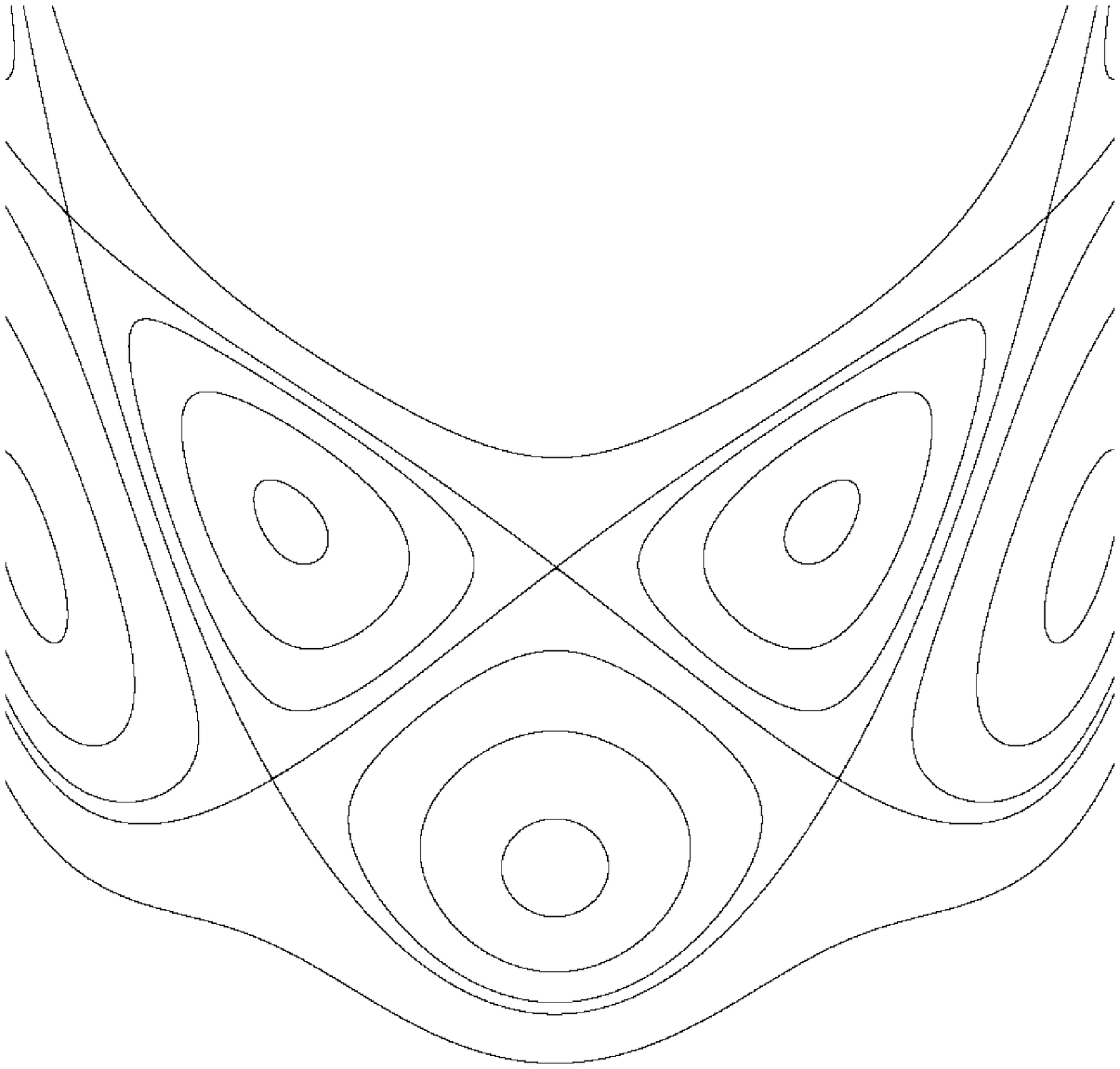}
\epsfxsize=4.6cm\epsfbox{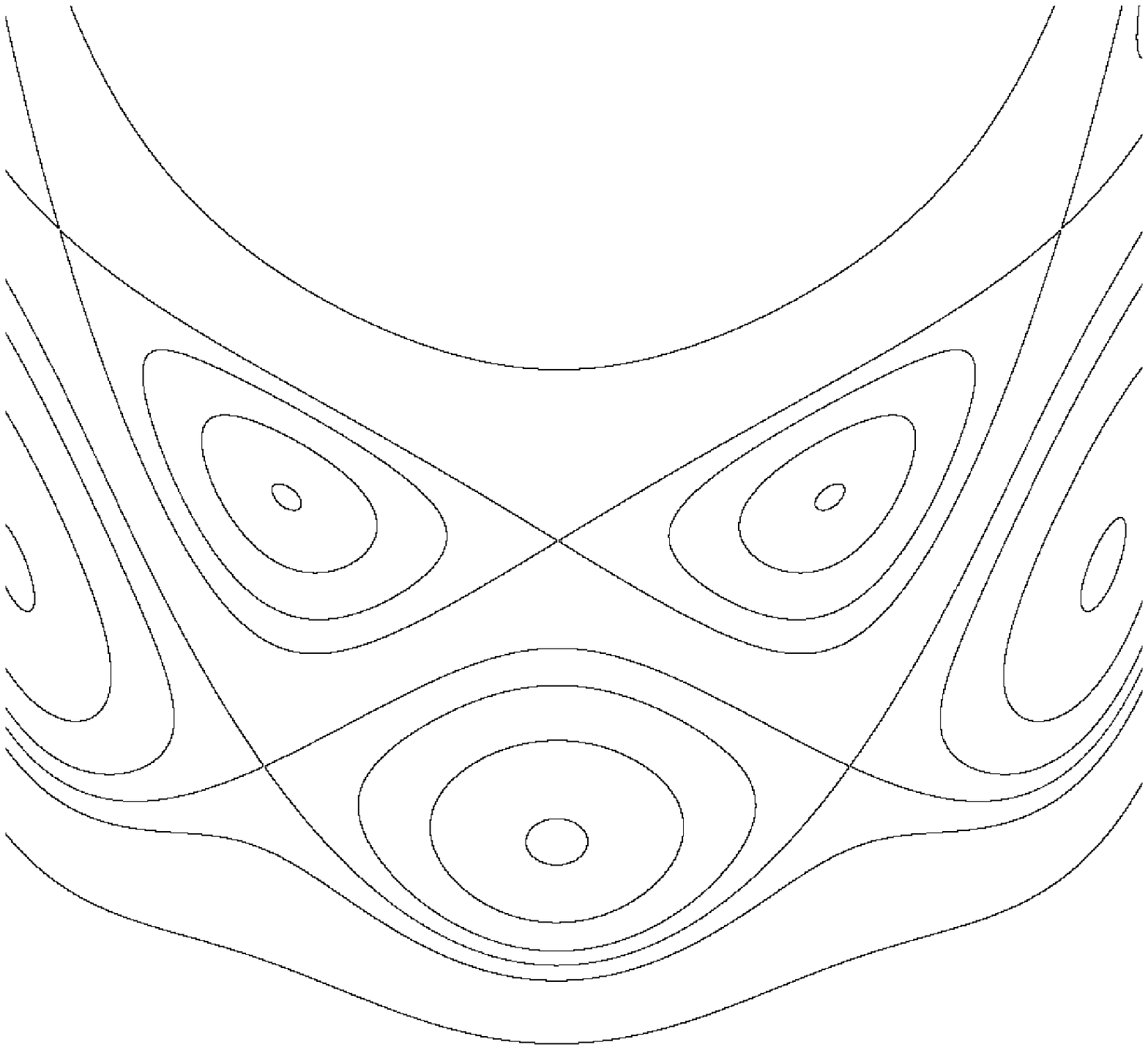}
\epsfxsize=4.6cm\epsfbox{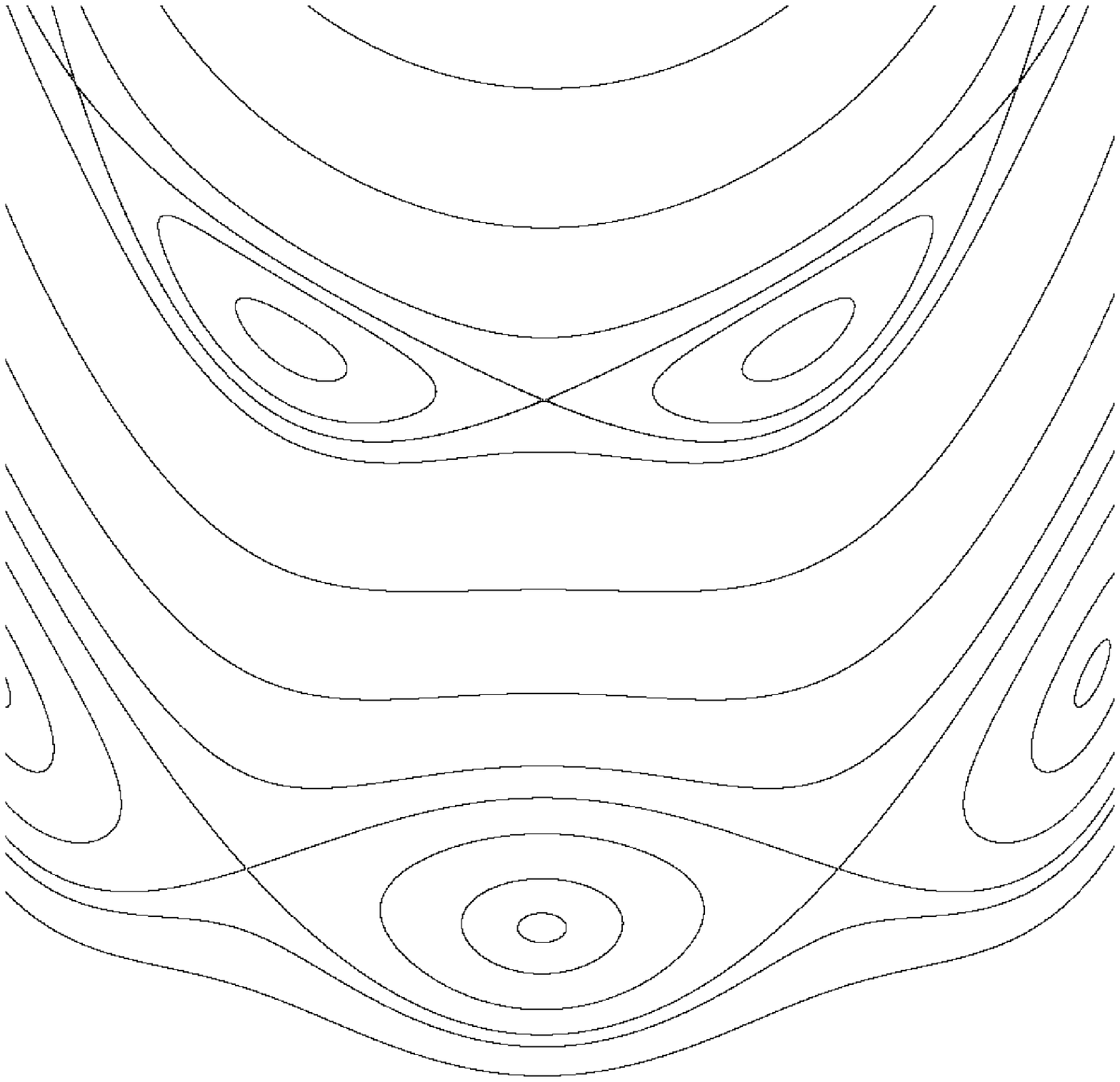}
}
}{ht}
{Phase portraits of the \hen map near $\omega_{3}$ 
where the $3/10$ reconnection bifurcation occurs.
$k=0.70638, 0.70639, 0.70640, 0.70640121, 0.70642, 0.70650$}
{fig:henonMap}{9cm}

\begin{table}[h]
	\centering
	\caption{Bifurcation Points for the \hen map}
	\begin{tabular}{c|l|l|l}
		Bifurcation & label & $\omega$ & $k$ \\
		\hline
		twistless       & $\omega_{0}$ & 0.2902153    & $\frac{9}{16}$  \\
		\hline
		sn($3/10$)      & $\omega_{1}$ & 0.2995432    & 0.7063832  \\
		\hline
		sn($3/10$)      & $\omega_{2}$ & 0.2995438    & 0.7063926  \\
		\hline
		reconnection    & $\omega_{3}$ & 0.2995444    & 0.70640121 \\
		\hline
		decupling       & $\omega_{4}$ &$\frac{3}{10}$& 0.7135255  \\
		\hline
		sn($1/3$)       & $\omega_{5}$ &0.3179717     & 1  \\
		\hline
		tripling        & $\omega_{6}$ &$\frac13$     & $\frac54$  \\
	\end{tabular}
	\label{tbl:henbifs}
\end{table}

\Epsfig{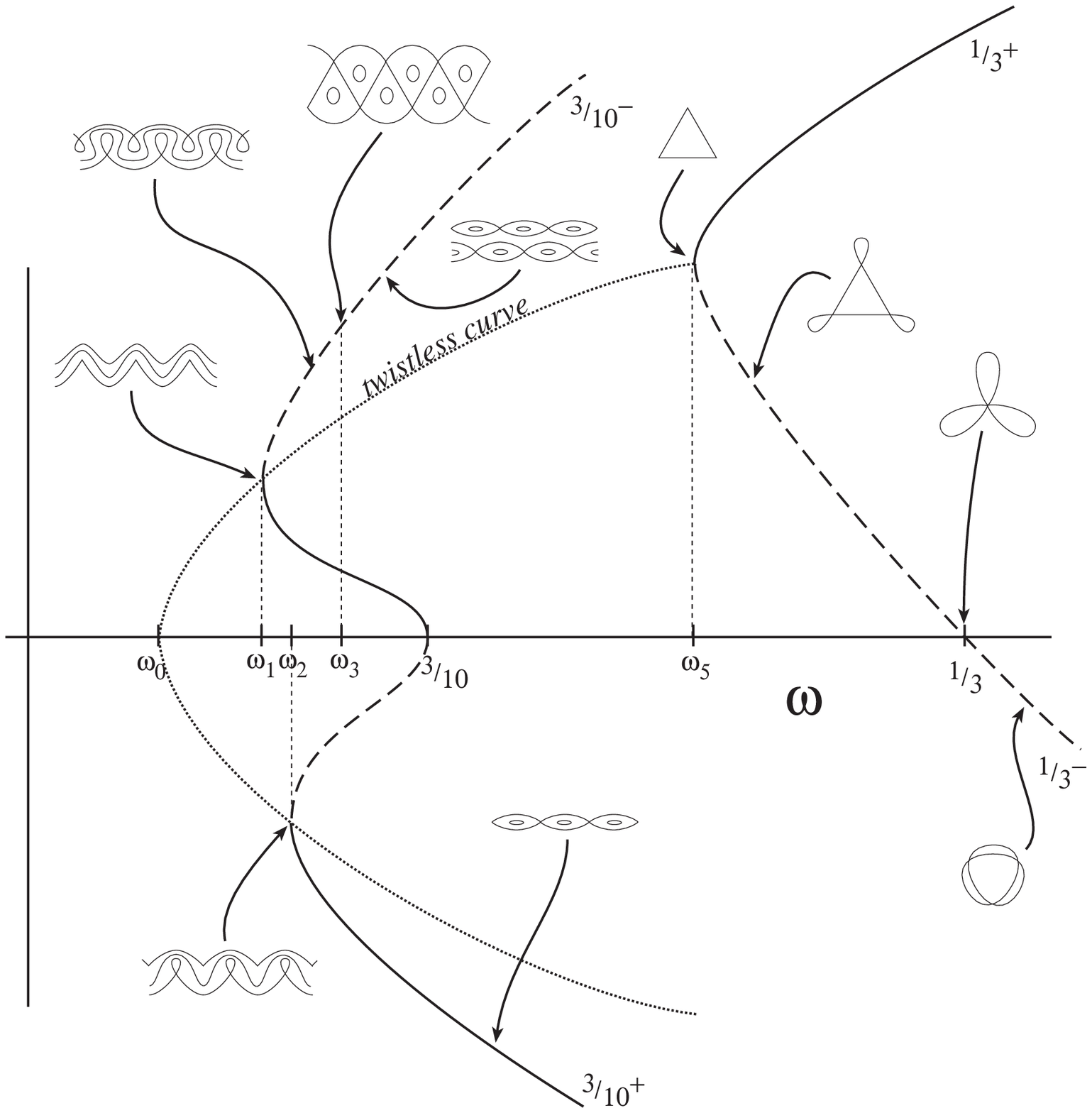}{ht}
{Sketch of the $3/10$ and $1/3$ bifurcations for the \hen map as a 
function of $\omega$. Representative phase portraits are also shown.
See Table~\ref{tbl:henbifs} for bifurcation values.}
{fig:bifwithten}{9cm}

\subsection{Normal form for a quartic map}\label{sec:quarticMap}
To study codimension two bifurcations, we need a map with two 
essential parameters. Since the quadratic map has only one (the 
rotation number), we turn to a higher order polynomial map which we take to be the 
composition of a rotation and a shear
\[
	\left({\begin{array}{l} x\\ y \end{array}}\right) 
	     \mapsto  {R}_{2 \pi \omega }
	              \left({\begin{array}{l} x\\ y + x^{2} +  c x^{3}+ d x^{4}\end{array}}\right)
\]
where $R_{\varphi}$ is a counterclockwise rotation by the angle 
$\varphi$.  This map has $3$ essential parameters, $\omega, c$, and $d$. 
According to our general theory, we expect that with a choice of one 
parameter, $\omega$, or $c$ we can set $\tau_{0} = 0$. In fact, 
with the choice of $c$, we can set $\omega_{0}$ to be any value we 
choose. Finally, with the choice of $d$  we can set $\tau_{1}=0$, as well.

Transforming this map into Birkhoff normal form, we obtain 
\Eq{eqn:bFormMap}, providing the nonresonance conditions are satisfied 
as usual.  The twists are
\begin{eqnarray*}
\pi\tau_0 &=& \frac{3c}{8} + \frac{5-3t^{2}}{8t(t^{2}-3)}\\
\pi\tau_1 &=&  \frac {-705+2055t^{2}-2706t^{4}+2038t^{
	       6}-637t^{8}+51t^{10}}{512t^{3}\left (t^{2}-3\right )^{3}
	       \left (t^{2}-1\right )} \\
  	   & &+ \frac{3c}{256t(t^{2}-1)} \left[ \frac{-225+556t^{2}-534t^{4}+204t^{6}-17t^{8}}
	      {t(t^{2}-3)^{2}} + \frac{c}{2} (17-38t^{2}+17t^{4})\right ]\\
       & &+  \frac{3d}{16} \frac{7-5t^{2}}{t(t^{2}-3)}
\end{eqnarray*}
For the cubic map with $d=0$ we can make $\tau_0=0$ at any
$\omega_0 \ne 1/3$ with an appropriate choice of $c$.
For example, we can set $\omega_{0} = 1/5$ or $2/5$, and then 
choose 
\[
	c = \frac16 \sqrt{10\pm 2\sqrt{5}}.
\]
When $\tau_{0}=0$ at a fifth order resonance, 
\Th{thm:instability} in the Appendix implies that the fixed point is unstable 
at the bifurcation point.  
The corresponding map with $\omega \approx 1/5$ is shown in \Fig{fig:cubicMap}.
The instability is difficult to see numerically because the size of the 
nonlinear terms is so small near the fixed point, though in the figure 
it is clear that the unstable motion does approach the origin.
It is interesting to note that the instability is much more 
pronounced for $\omega = 1/5$ than for $\omega = 2/5$.

The second twist $\tau_1$ can be used in two different ways.
For the cubic map with $d=0$ we can show that 
if $\tau_0 = 0$ then always $\tau_1 \ne 0$, so that
the only exceptions to stability can be the third through sixth
order resonances. The value of $c$ that makes $\tau_0 = 0$ at $t=t_0$
and the corresponding value of $\tau_1$ are
\[
	c = \frac{5-3t_0^2}{3t_0(t_0^2-3)}, \quad
	\pi\tau_1 = \frac{105 - 305 t_0^2 + 353 t_0^4 - 167 t_0^6 + 30 t_0^8}
			{48t_0^3(t_0^2-3)^3(t_0^2-1)}.
\]
It is easy to see that 
the polynomial in the numerator of $\tau_1$ is positive, so 
that $\tau_1$ is always nonzero if $\tau_0$ vanishes for the cubic map.

For the quartic map with $d \ne 0$ we can make both twists
vanish at arbitrary $\omega_0$, in particular at seventh order resonance.
In general
\[
	c = \frac{5-3t_0^2}{3t_0(t_0^2-3)}, \quad
	d = \frac{105 - 305 t_0^2 + 353 t_0^4 - 167 t_0^6 + 30 t_0^8}
			{9t_0^2(t_0^2-3)^2(t_0^2-1)(5t^2-7)},
\]
and with $t_0 = \tan(\pi/7)$ we obtain
$c \approx 1.0763012774$ and $d \approx .7144291292$.
By \Th{thm:instability} the fixed point is unstable for these parameter
values.
Note that if we let $c$ and $d$ be the above functions of $\omega$
we obtain the very special one parameter family of quartic maps for
which the two first twists vanish for every frequency.

\rem{
With a choice of $c$, we can set $\tau_0=0$ for any $\omega_{0} \ne 
1/3$. 
In addition, when $\omega 
\ne 1/4$, we can choose $d$ so that $\tau_{1} = 0$.  For example, we 
can set $\omega_{0} = 1/5$ or $2/5$, by choosing
\begin{eqnarray*}
    c = \frac13 \sqrt{ \frac{10}{5 \mp \sqrt{5}} }  \quad , \quad
    d =  -\frac{5}{18} \frac{2\sqrt{5} \pm 11}{ \sqrt{5} \mp 15}  \;,
\end{eqnarray*}
respectively. 
}


\EpsfigStuff{
\centerline{
\epsfxsize=5cm
	\epsfbox{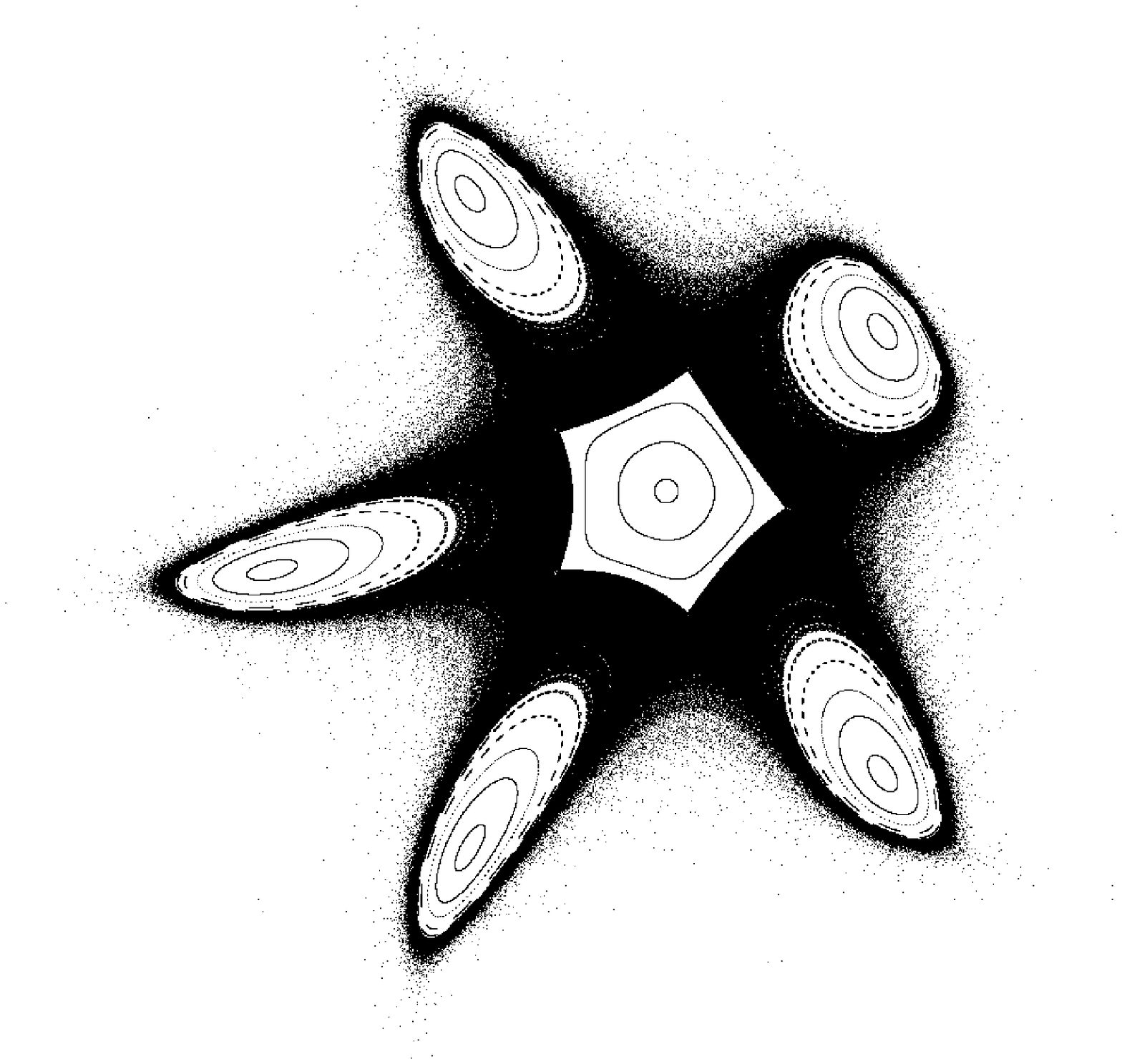}
\epsfxsize=5cm
	\epsfbox{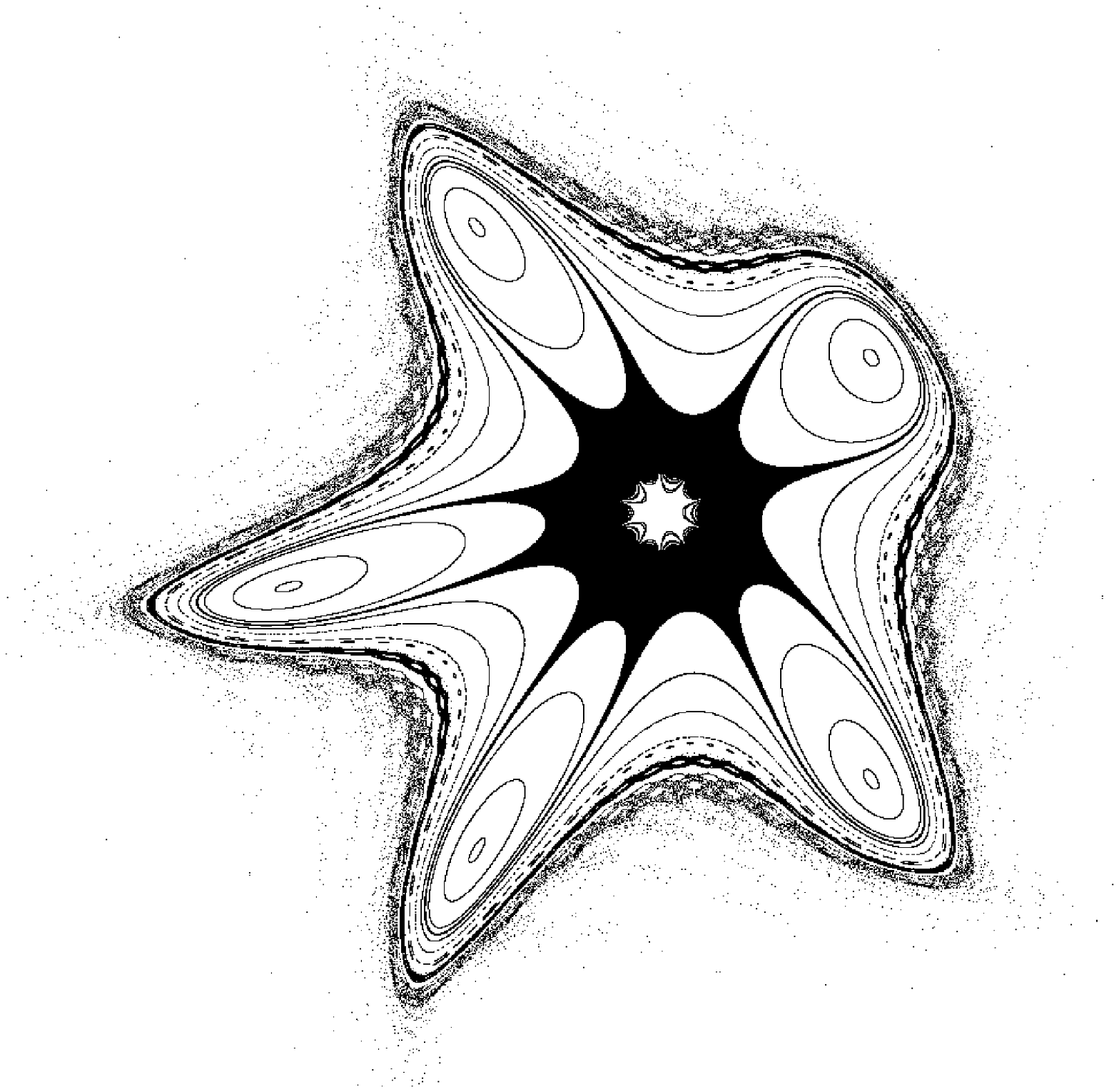}
\epsfxsize=5cm
	\epsfbox{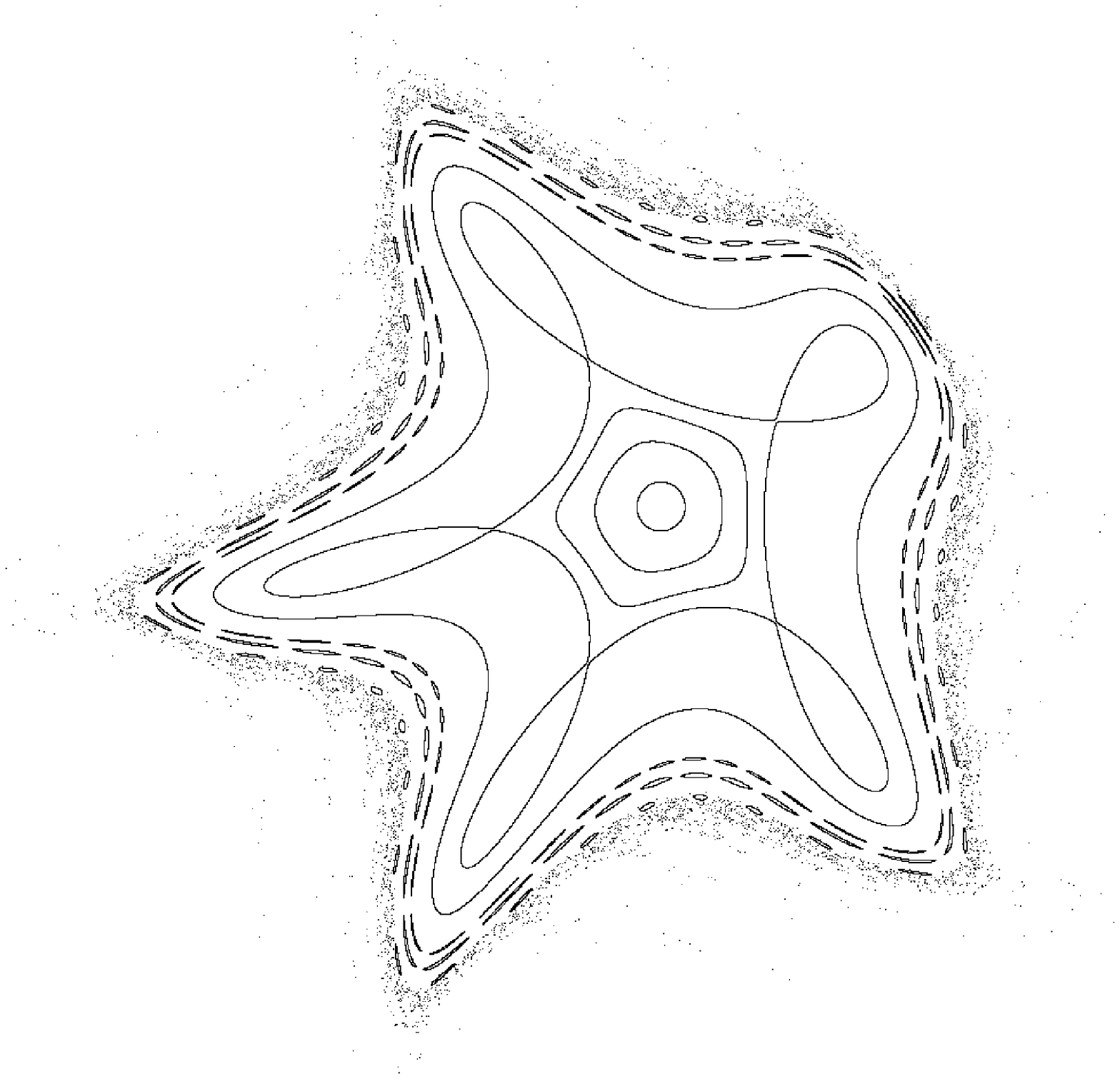}
}
}{ht}
{Phase portrait of the cubic map (d=0) with $\tau_{0}=0$ at $\omega = 
1/5-\epsilon,1/5,1/5+\epsilon$, $\epsilon = 0.0004$.  
At $\omega = 1/5$ the fixed point is shown to be unstable.}
{fig:cubicMap}{9cm}


\section{Conclusion}

By analyzing the normal form for Hamiltonian flows and area-preserving 
maps in the neighborhood of the tripling bifurcation we have shown 
that there exists a bifurcation creating a twistless torus.  Thus in 
the presence of a tripling bifurcation, the non-degeneracy condition of 
the KAM theorem is violated on a one parameter family of tori in phase 
space in the neighborhood of the bifurcation.  Moreover, 
in the nonintegrable case, reconnection 
bifurcations occur for orbits that collide at the twistless torus, 
such as the $3/10$ orbits of the \hen map in \Fig{fig:bifwithten}.

Using the notation of \Fig{fig:bifwithten}, we say that a twistless 
bifurcation occurs at $\omega = \omega_{0}$.  The pair of period three 
orbits is created in a saddle-center bifurcation at $\omega_{5}$, and 
the tripling occurs at $\omega_6 = 1/3$.  For the case of the 
resonant Hamiltonian normal form, \Eq{eqn:epszero} implies that the 
relative positions of these bifurcations have a simple ratio
\[ 
     \frac{\omega_{0}-1/3}{\omega_{5}-1/3} = \frac83 \; .
\]
Since the Hamiltonian resonant normal form is integrable, we can 
follow the twistless torus; it exists when the rotation number of the 
central periodic orbit is in the interval $\omega_{0}$ to $\omega_5$.  
Moreover, the rotation number of the twistless curve itself, 
$\Omega_{0}$, takes all of the values in the interval from $\omega_0$ to
$1/3$.  For every 
rotation number in this interval there will generically be a 
reconnection bifurcation for the full two degree of freedom 
Hamiltonian. Hence this type of bifurcation occurs generically in 
nonintegrable Hamiltonian systems.

Similarly any area-preserving map that has a tripling bifurcation of 
an elliptic fixed point and that satisfies the nondegeneracy condition 
that the third order resonant coefficient, $a_{02}$ does not vanish, 
has a twistless bifurcation.  The dependence of the twist on rotation 
number is more complicated for this case than for the Hamiltonian, 
nevertheless, the rotation number of the fixed point unfolds the 
bifurcation.  Moreover, there is always an $\omega \in [1/6,1/2)$ for 
which the resonant normal form has a twistless bifurcation.

As an example, the \hen map has a twistless bifurcation at $k=9/16$, 
and the twistless curve moves outward as $k$ increases, causing 
reconnection bifurcations.  We examined in particular the bifurcations 
of the $3/10$ orbits.  The vanishing of twist(s), as we showed for a 
cubic/quartic area-preserving map, can also lead to the instability of the 
fixed point at quintupling/septupling bifurcations.

The whole scenario is similar for the quadrupling bifurcation, when the 
resonant term dominates the twist term in magnitude (see the 
Appendix).  In this case a pair of period $4$ orbits are created in a 
saddle-center bifurcation. and a twistless torus is created at a 
cusp in the bifurcation diagram similar to that in \Fig{fig:bifHam3}.  
However, in this case the twistless curve does not collide with the 
fixed point, and a simple prediction about what rotation numbers must 
pass through the twistless curve is not possible.

It would be interesting to generalize these results to higher 
dimensional systems.


\section{Appendix:\\Instability of resonant maps without twist}

In this section we give sufficient conditions for which a map of the 
plane with eigenvalues that are $n^{\rm th}$ roots of unity 
has an unstable fixed point. Here the condition of area preservation 
is not necessary. Examples of this behavior are well known 
\cite{SM71}, however, we have not found an explicit proof of 
instability for the case where the resonant
term has the lowest possible order.

To study this problem, we put the map in the normal form \Eq{eqn:rFormMap}.
We will show that the fixed point is generically unstable whenever all of the twist terms 
vanish that are of lower order than $O(n-1)$, the order of the resonant 
term.  When $n$ is even, there is a twist term of the same order as 
the resonant term. In this case the $O(n-1)$ twist term need not vanish, 
but its magnitude must be dominated by the resonant term.

\begin{teo}\label{thm:instability}
	Let $f$ be a $C^{n}$ map of the plane with a fixed point at the 
	origin.  Suppose that the multipliers of the fixed point are 
	$n^{\rm th}$ roots of unity, and that when $f$ is put in normal 
	form, the low order twist coefficients vanish:
	\[
	     a_{j,j-1} = 0 \quad, \quad  1<j<\frac{n}{2} \,,
	\]
	    but the lowest order resonant term is nonzero
	\[
	     a_{0,n-1} \ne 0 \,.
	\]
	If $n$ is even, assume in addition that
	\[
	     |a_{n/2,n/2-1}| < |a_{0,n-1}| \,.
	\]
	Then the origin is unstable.
\end{teo}

Note that when $n=3$, the assumed form is generic, but for $n=4$, the 
first twist, $a_{21}$, is assumed to be small, and when $n=5$ it is 
assumed to vanish.

The instability at the (generic) tripling bifurcation is well known.  
Our calculation shows that a similar phenomenon occurs for the 
quintupling bifurcation: the fixed point becomes unstable 
if the first twist vanishes.  As we saw in \S{sec:quarticMap} this is 
a generic bifurcation in a two parameter area-preserving family, 
one parameter to adjust the 
frequency and the second one to make the first twist vanish.

\proof
When $n$ is odd the assumed normal form is 
\[
	z' = \lambda( z + \alpha \bar z^{n-1}) + O(n) \;.
\]
We adapt the proof from Siegel and Moser \cite{SM71}.  Begin by 
scaling $z$ using the transformation $z \to zs$ where $s$ determined 
by $\bar s^{n-1}/s = (n\alpha)^{-1}$.  Then, using the fact that 
$\lambda^{n}=1$, the $n^{\rm th}$ power of $z'$ becomes
\[
    {(z')}^{n} = z^n + z^{n-1} \bar z^{n-1} + O(2n-1) \;.
\]
Introducing new coordinates $Z$ by
\[
	z^{n} = Z = X + iY = R \exp{i\Phi} \;,
\]
gives
\begin{eqnarray*}
	Z' &=& Z + R^{2-2/n} \left(1+\eta(Z,\bar Z) R^{1/n}\right) \;,
\end{eqnarray*}
where the higher order terms have been represented by a factor 
$\eta$, which, since $f$ is $C^{n}$, is bounded as $R \to 0$.
The purpose of the scaling to eliminate $\alpha$ is that the 
real part of this map now becomes
\[
	X' = X + R^{2-2/n}\left(1+\Re(\eta) R^{1/n}\right) \;.
\]
Since $\eta$ is bounded, there is an $R_{0}$ such that for all 
$R<R_{0}$, we have $|\Re(\eta) R^{1/n}| < 1/2$.  Thus whenever $R<R_{0}$, $X' 
> X + \frac12 R^{2-2/n}$, and so $X$ is monotone increasing in each 
step of the map.  It is easy to see that this implies that the fixed point 
is unstable.

The case of even $n$ is slightly harder because there can be a twist term 
of the same order:
\[
	z' = \lambda\left( z + \alpha \zb^{n-1} + \beta z|z|^{(n-2)}
			+ O(2n-1) \right).
\]
If $\beta$ happens to be zero the above proof works.  When
$\beta$ is nonzero the stability of the fixed point is determined by the 
relative size of the resonant term and the twist term.  Again we scale 
away $\alpha$ and the $n^{\rm th}$ power of the new map is
\[
{(z')}^{n} = z^n + (z\zb)^{n-1} + \tilde \beta z^n |z|^{(n-2)} + O(2n-1),
\]
where $\tilde \beta = \beta/|\alpha|$.
In the new coordinates this becomes
\[
	Z' = Z + R^{2-2/n}(1 + \tilde \beta \exp(i\Phi) + \eta R^{1/n}),
\]
and the map for the real part is
\[
    X' = X + R^{2-2/n}\left(1 + \Re(\tilde\beta \exp(i\Phi)) + \Re(\eta)R^{1/n}
	\right).
\]
So if $1 + \Re(\tilde\beta\exp(i\Phi)) > 0$ for all $\Phi$ the
argument of the previous case applies.  Therefore the instability 
criterion is $|\beta|/|\alpha| < 1$:  if the ``twist'' 
term is smaller than the resonant term the mapping is 
unstable.  The well known example of this type is the 
quadrupling bifurcation.  Our calculation shows that a similar 
codimension two bifurcation occurs for the 
area preserving sextupling bifurcation.
\qed

\bibliographystyle{alpha}
\bibliography{notwist3}

\begin{thebibliography}{dCNGM96}

\bibitem[Arn78]{Arnold78}
V.I. Arnold.
\newblock {\em Mathematical Methods of Classical Mechanics}.
\newblock Springer, New York, 1978.

\bibitem[dA90]{Ozorio88}
A.M.~Ozorio de~Almeida.
\newblock {\em Hamiltonian Systems: Chaos and Quantization}.
\newblock Cambridge University Press, Cambridge, 1990.

\bibitem[dCNGM96]{CGM96}
D.~del Castillo-Negrette, J.M. Greene, and P.J. Morrison.
\newblock Area preserving nontwist maps: Periodic orbits and transition to
  chaos.
\newblock {\em Physica D}, 91(1):1--23, 1996.

\bibitem[DdlL98]{DL98}
A.~Delshams and R.~de~la Llave.
\newblock {KAM} theory and a partial justification of greene's criterion for
  non-twist maps.
\newblock {\em preprint}, 1998.

\bibitem[H{\'e}n69]{Henon69}
M.~H{\'e}non.
\newblock Numerical study of quadratic area-preserving mappings.
\newblock {\em Quart. Appl. Math.}, 27:291--312, 1969.

\bibitem[HH84]{HowHoh84}
J.E. Howard and S.M. Hohs.
\newblock Stochasticity and reconnection in hamiltonian systems.
\newblock {\em Physical Review A}, 29:418, 1984.

\bibitem[HH95]{HowHum95}
J.~E. Howard and J.~Humpherys.
\newblock Nonmonotonic twist maps.
\newblock {\em Physica D}, 80(3):256--276, 1995.

\bibitem[Mei92]{Meiss92}
J.D. Meiss.
\newblock Symplectic maps, variational principles, and transport.
\newblock {\em Reviews of Modern Physics}, 64(3):795--848, 1992.

\bibitem[MH92]{MH92}
K.R. Meyer and G.R. Hall.
\newblock {\em Introduction to the Theory of Hamiltonian Systems}, volume~90 of
  {\em Applied Mathematical Sciences}.
\newblock Springer-Verlag, New York, 1992.

\bibitem[Mos94]{Moser94}
J.K. Moser.
\newblock On quadratic symplectic mappings.
\newblock {\em Math. Zeitschrift}, 216:417--430, 1994.

\bibitem[Sch98]{Schomerus}
H.~Schomerus.
\newblock Periodic orbits near bifurcations of codimension two: classical
  mechanics, semiclassics and stokes transitions.
\newblock {\em J. Phys. A}, 31(18):4167--4196, 1998.

\bibitem[SDM98]{SDM98}
D.~Sterling, H.R. Dullin, and J.D. Meiss.
\newblock Homoclinic bifurcations for the {H}\'enon map.
\newblock {\em submitted to Physica D}, 1998.

\bibitem[Sim99]{Simo98}
C.~Sim{\'o}.
\newblock Invariant curves of analytic perturbed nontwist area preserving maps.
\newblock {\em Regular {\&} Chaotic Dynamics}, 1999.
\newblock to appear.

\bibitem[SM71]{SM71}
C.L. Siegel and J.K. Moser.
\newblock {\em Lectures on Celestial Mechanics}.
\newblock Classics in Mathematics. Springer-Verlag, New York, 1971.

\bibitem[ST98]{SimTre98}
C.~Sim{\'o} and D.~Treschev.
\newblock Evolution of the {``last''} invariant curve in a family of area
  preserving maps.
\newblock {\em preprint}, 1998.

\bibitem[VV90]{WVCP88}
J.~P. Vanderweele and T.~P. Valkering.
\newblock The birth process of periodic orbits in nontwist maps.
\newblock {\em Physica A}, 169(1):42--72, 1990.

\end{thebibliography}

\end{document}